\pdfoutput=1
\documentclass[a4paper]{article}
\usepackage{graphicx,epsfig,color}
\usepackage{jheppub}
\usepackage[english]{babel}
\usepackage{amssymb,amsfonts,amsmath}
\usepackage{verbatim}
\usepackage{physics}
\usepackage[dvipsnames]{xcolor}
\usepackage{bbold,bbm,bm}
\usepackage[normalem]{ulem}
\newcommand{\be}{\begin{equation}} \newcommand{\ee}{\end{equation}}
\newcommand{\bea}{\begin{eqnarray}} \newcommand{\eea}{\end{eqnarray}}

\usepackage{cancel}
\usepackage{float}
\usepackage{booktabs}

\begin{document}

\begin{flushright}
HIP-2026-22/TH
\end{flushright}

\title{Effective scalar--fluid description of thermal first-order phase transitions and their spinodal instabilities}

\author[a]{Alessio Caddeo,}
\affiliation[a]{Institute for Theoretical Physics and Astrophysics and W\"urzburg-Dresden
Cluster of Excellence ctd.qmat, Julius-Maximilians-Universit\"at W\"urzburg, 
Am Hubland, DE-97074 W\"urzburg, Germany}
\emailAdd{alessio.caddeo@uni-wuerzburg.de}

\author[b,c]{Oscar Henriksson,}
\affiliation[b]{Department of Natural and Health Sciences, {\AA}bo Akademi University, Henrikinkatu 2, FI-20500 Turku, Finland}
\affiliation[c]{Helsinki Institute of Physics, P.O.~Box 64, FI-00014 University of Helsinki, Finland}
\emailAdd{oscar.henriksson@abo.fi}

\author[d,e]{Carlos Hoyos,}
\affiliation[d]{Department of Physics, Universidad de Oviedo, c/ Leopoldo Calvo Sotelo 18, ES-33007 Oviedo, Spain}
\affiliation[e]{Instituto de Ciencias y Tecnolog\'{\i}as Espaciales de Asturias, c/ Independencia 13, ES-33004 Oviedo, Spain}
\emailAdd{hoyoscarlos@uniovi.es}

\author[f]{Mikel Sanchez-Garitaonandia,}
\affiliation[f]{School of Mathematics \& Statistics, University College Dublin, Belfield, Dublin 4, D04 V1W8, Ireland}
\emailAdd{mikel.sanchezgaritaonandia@ucd.ie}

\author[g,h]{Pedro Taranc\'on-\'Alvarez}
\affiliation[g]{Departament de F\'\i sica Qu\`antica i Astrof\'\i sica,  Universitat de Barcelona, Mart\'\i\  i Franqu\`es 1, 
\mbox{ES-08028}, Barcelona, Spain}
\affiliation[h]{Institut de Ci\`encies del Cosmos (ICC),  Universitat de Barcelona, Mart\'\i\  i Franqu\`es 1, 
\mbox{ES-08028}, Barcelona, Spain}
\emailAdd{pedro.tarancon@icc.ub.edu}

\abstract{We formulate an improved scalar--fluid effective description of first-order phase transitions in relativistic theories, and test it against a microscopic calculation in a holographic model. The effective theory shows that the nature of the spinodal instability is linked to thermodynamic stability, determined by the sign of the specific heat. When the specific heat is positive, the sound modes are stable and there is a non-hydrodynamic instability. When it is negative, a sound mode becomes unstable instead. The holographic calculation confirms the validity of the effective description at long wavelengths, improving on the purely hydrodynamic description. We also identify a friction coefficient that governs the bubble wall velocity and can be extracted from a holographic correlator.}

\maketitle
\flushbottom
\setcounter{page}{2}

\newpage

\section{Introduction}\label{sec:intro}

Phase transitions can take place in systems that span a huge range of scales, from nuclear physics to cosmology. A first-order phase transition (FOPT) taking place in the early universe can leave an observable signal on the cosmological gravitational wave background that could be detected in the near future. This would be a smoking gun for physics beyond the Standard Model, which lacks first-order transitions in the region of the phase diagram explored by the cosmological evolution (see \cite{Caprini:2019egz,Hindmarsh:2020hop} for reviews). 

A FOPT takes place when at some value of the temperature, or other thermodynamic parameter, the equilibrium state in which the system is found becomes metastable, and can evolve to another globally stable equilibrium state. The transition typically proceeds through bubble nucleation of the stable phase inside the metastable phase. These bubbles appear at random with some probability and, above some critical size, expand and collide, eventually leaving only the homogeneous stable phase or a phase-separated state. 

However, in some physical systems fluctuations are suppressed, and the resulting nucleation rate can be very small. This can result in strong supercooling, with the system potentially unable to complete the transition through bubble nucleation. Such scenarios have been realized experimentally in some materials \cite{Bar_2018,Kundu_2020}. If the system evolves slowly enough to remain near equilibrium, it may eventually reach a point where the metastable phase terminates. Both the high- and low-temperature phases of an FOPT typically end at such \emph{spinodal points}, which are connected by a branch of linearly unstable states, the spinodal branch (see e.g. Fig.~\ref{fig:s-of-T} below). These states can be unstable in two distinct ways. If they are thermodynamically unstable with negative specific heat, the sound modes are unstable and drive the transition. If the specific heat is positive, the sound modes are stable, and the instability must instead reside in a non-hydrodynamic mode.

In this work, we propose an effective field theory (EFT) of a fluid coupled to a scalar order parameter, extending previous proposals introduced to describe the evolution of nucleated bubbles in cosmological phase transitions \cite{Ignatius:1993qn,Moore:1995si}. Within this effective description, we then derive a link between thermodynamic stability and the type of dynamical instability in the unstable phase.

We test the validity of the EFT at the linear level by comparing with a microscopic calculation for strongly coupled theories with holographic duals. The holographic duality, or AdS/CFT correspondence \cite{Maldacena:1997re,Witten:1998qj,Gubser:1998bc} has previously been used to determine the bubble wall velocity \cite{Bea:2021zsu,Bigazzi:2021ucw,Henriksson:2021zei,Bea:2022mfb,Bea:2024bls,Henriksson:2024hsm}, estimate the gravitational wave spectrum \cite{Ares:2020lbt,Bea:2021zol, Bigazzi:2020avc}, and to find an effective action for the order parameter \cite{Ares:2021nap,Ares:2021ntv}. Spherical critical bubbles have recently been constructed in \cite{Henriksson:2025vci,Hoyos:2026eoq}, and match expectations from the effective theory. Analogous comparisons of effective theories and microscopic holographic models with a symmetry-breaking second-order phase transition have also been studied in \cite{Donos:2022qao, Donos:2025jxb, Donos:2025igh}.

The duality is holographic in the sense that the strongly coupled theory has a dual which lives in a space with one additional spatial dimension. The additional dimension roughly encodes energy or length scales. The strongly coupled theory is a quantum field theory (QFT) and the dual is a weakly coupled gravitational theory, which becomes classical in the limit in which the QFT has a large number of degrees of freedom (referred to as the large-$N$ limit). This limit suppresses fluctuations, and particularly the nucleation rate. This implies that holographic models typically exhibit strong supercooling \cite{Ares:2021nap}, and may even be completely unable to complete the transition through bubble nucleation.

The possibility of a complete suppression of bubble nucleation in holographic models has been considered in two scenarios. In \cite{Bea:2021zol}, it was argued that there can be a quasistatic decrease of the energy density such that the system remains close to equilibrium in the metastable state until it enters an unstable state. This phase has unstable sound modes that are identified with a spinodal instability that would introduce density inhomogeneities driving the system to a phase-separated state. In contrast, in \cite{Caddeo:2024lfk} it was found that, for a slowly changing temperature, the metastable phase exhibits critical behavior and the quasistatic evolution breaks down before reaching the unstable phase. Furthermore, in this setup the unstable phase does not have an instability related to sound, but a different type of non-hydrodynamic instability inducing a homogeneous change of the order parameter, in addition to spinodal decomposition.

The main difference between the two cases just described is the size of the sector undergoing the phase transition. In the case where the sound is unstable, the phase transition takes place in a sector with a number of degrees of freedom comparable to that of the whole system. In this situation the temperature may not be a good variable to keep track of the evolution, as it is a concept that relies on equilibrium thermodynamics. The energy density on the other hand is a well-defined observable in any equilibrium or non-equilibrium state. In the other case, with stable sound modes, the phase transition takes place in a sector with a number of degrees of freedom that is much smaller than that of the whole system. This means that this sector can be treated as a probe, and the rest of the system acts as a thermal bath with a well-defined temperature, at least in some approximate sense. This situation is actually analogous to that of weakly coupled models used in cosmological transitions. In practice, the thermal bath is described in the gravity dual by the metric, and the sector undergoing the transition by a scalar field. In the first case the scalar field fully backreacts on the metric, while in the second the scalar field is a probe and the backreaction is completely neglected.

The difference between the number of degrees of freedom can be translated into the thermodynamic stability of the unstable phase. The sound mode instability can be identified with having a negative specific heat $c_v<0$, which gives an imaginary speed of sound $c_s$ through the usual formula $c_s^2=s/c_v$, with $s$ the entropy density. In the second case we discussed, the value of the specific heat is determined to leading order by the thermal bath and remains positive in the unstable phase, which explains the absence of unstable sound modes. The instability is restricted to the sector undergoing the phase transition, which is basically captured by the order parameter. It is actually possible to find models where the specific heat changes between positive and negative values within the same unstable phase. Fig.~\ref{fig:s-of-T} shows two possible cases, one where the specific heat changes sign in the unstable branch and another where it is negative throughout. In both cases, the sign of the specific heat can be read directly from the slope of the curves.

An obvious question is whether the dynamical instability also changes between hydrodynamic and non-hydrodynamic accordingly. As we will show, this is indeed the case, and this behavior is correctly captured by the effective fluid plus scalar description.

The paper is organized as follows: in section~\ref{sec:scalar} we introduce the effective theory of the coupled scalar and fluid and discuss its thermodynamics and spectrum of linear fluctuations around equilibrium. We also explain the relation between backreaction and thermodynamic stability. In section~\ref{sec:holog} we introduce the microscopic theory as a holographic model, derive its thermodynamics and study the phase diagram with changing backreaction. Next, in section~\ref{sec:qnms} we compute the spectrum of quasinormal modes in the holographic theory, fit the parameters of the effective theory at low momentum, and test the validity of the effective theory by comparing dispersion relations at finite momentum with the quasinormal modes. Finally, in section~\ref{sec:disc}, we summarize and discuss our results and possible future directions.

\section{Scalar field coupled to a fluid}\label{sec:scalar}

The description of bubble dynamics in a cosmological phase transition is typically discussed in the context of the electroweak transition. A review can be found in \cite{Hindmarsh:2020hop}, based on the original work \cite{Ignatius:1993qn,Moore:1995si}. The system consists of a scalar field, that is treated semiclassically, and a thermal bath of weakly interacting particles, that is described using kinetic theory. The effective theory is that of a classical scalar field coupled to a fluid. The interaction between the two emerges from the dependence of the masses of the particles on the field, which produces an effective force on the particles when the gradients of the field are non-zero. Since the total energy and momentum are conserved, there is a backreaction on the scalar that modifies its equation of motion. Other effects, such as momentum exchange with the scalar during particle collisions, are neglected when the scalar changes over a region much larger than the mean free path of the particles.

Fluid systems coupled to scalar degrees of freedom have been obtained using the fluid/gravity duality~\cite{Bhattacharyya:2008ji,Blake:2015epa,Blake:2015hxa}, although there the scalars are prescribed sources that break translation invariance explicitly, rather than dynamical order parameters. A study of those cases from an effective field theory approach was done in~\cite{Gouteraux:2023uff}. 

Superfluid hydrodynamics also includes a dynamical scalar field whose gradient, the superfluid velocity, appears in the constitutive relations. Contrary to our case, this gradient is a genuine thermodynamic variable that enters the free energy. Holography can also be used there to extract the transport coefficients and to study the interplay between thermodynamics and dynamical instabilities~\cite{Gouteraux:2022qix,Arean:2023nnn}.

Let us briefly review the equations for the scalar and the fluid that arise from this picture. The conservation of the energy-momentum of the full system is
\begin{equation}
 0=  \partial_\mu T^{\mu\nu}=\partial_\mu \left(T_{\rm f}^{\mu\nu}+T_\phi^{\mu\nu}\right) \ ,
\end{equation}
with $T_{\rm f}^{\mu\nu}$ and $T_\phi^{\mu\nu}$ the energy-momentum tensors of the fluid and the scalar, respectively.

The energy-momentum of the fluid is taken to be the one of an ideal fluid
\begin{equation}
    T^{\mu\nu}_{\rm f}=(\varepsilon+p) u^\mu u^\nu+p \eta^{\mu\nu},
\end{equation}
where $\varepsilon$ is the energy density, $p$ is the pressure, $u^\mu$ the velocity ($u^\mu u_\mu=-1$) and $\eta_{\mu\nu}$ is the Minkowski metric in mostly plus signature. The pressure includes both the pressure of the thermal bath and the contributions from the scalar potential
\begin{equation}\label{eq:weakpressure}
    p=\frac{\pi^2}{90}g_{\rm eff} T^4-V_T(\phi),
\end{equation}
where $g_{\rm eff}$ measures the effective number of degrees of freedom in the thermal bath, and $V_T$ is the effective potential, including loop corrections from the particles coupled to the scalar.

The energy-momentum of the scalar, after removing the potential term contribution, is the one corresponding to the kinetic terms of a relativistic free field
\begin{equation}
    T^{\mu\nu}_\phi=\partial^\mu\phi\partial^\nu \phi-\frac{1}{2}\eta^{\mu\nu}\partial_{\alpha}\phi \partial^\alpha\phi
\end{equation}
The equation of motion for the scalar that is derived from this picture takes the form
\begin{equation}
    -\partial_\mu \partial^\mu \phi + \partial_\phi V_T=-\gamma u^\mu\partial_\mu \phi,
\end{equation}
where the term on the right-hand side is the effective force that is produced by the change in the mass of the particles. Projecting the conservation equation on the fluid velocity and using the equation of motion of the scalar, one arrives at
\begin{equation}
    0=-u_\nu \partial_\mu T^{\mu\nu}=\partial_\mu(\varepsilon u^\mu)+p \partial_\mu u^\mu -\partial_\phi V_T u^\mu\partial_\mu \phi-\gamma (u^\mu \partial_\mu \phi)^2.
\end{equation}
Using now the thermodynamic relations
\begin{equation}
    \varepsilon+p=Ts,\quad \partial_T\varepsilon=T\partial_T s, \quad \text{and} \quad \partial_\phi \varepsilon=T\partial_\phi s+\partial_\phi V_T,
\end{equation}
where $s$ denotes the entropy density, one obtains an equation for the entropy current which is consistent with the local form of the second law for $\gamma\geq 0$:
\begin{equation}
    T\partial_\mu(s u^\mu)=\gamma (u^\mu \partial_\mu \phi)^2.
\end{equation}
The terms proportional to $\gamma$ encode the information about the friction that the fluid exerts on the wall of an expanding bubble, and ultimately determine the bubble wall velocity. In later sections we will study how $\gamma$ enters in the dispersion relations of fluctuations around equilibrium, particularly concerning the spinodal instability in the unstable branch of the phase diagram. Here, we will first provide a generalization of this picture that does not assume Lorentz invariance of the kinetic terms in the effective action for the scalar, and that includes viscous terms in the energy-momentum tensor of the fluid, that could be of the same order as the gradients of the scalar in a derivative expansion.

\subsection{Generalization of the effective action and dissipative terms}\label{sec:effact}

We can combine the scalar field and the fluid in a single effective action that captures the non-dissipative part of the system. For the fluid, we follow \cite{Banerjee:2012iz,Jensen:2012jh}, where one introduces a nowhere vanishing time-like vector field $U^\mu$ coupled to a background metric $g_{\mu\nu}$, such that the local temperature and fluid velocity are given by
\begin{equation}
    T=\frac{T_0}{\sqrt{-U^2}},\quad u^\mu=\frac{U^\mu}{\sqrt{-U^2}},\quad U^2=g_{\mu\nu}U^\mu U^\nu,
\end{equation}
with $T_0$ a fixed scale. The action that captures ideal fluid hydrodynamics is simply
\begin{equation}
    S_{\rm f}=\int d^d x \sqrt{-g}P(T).
\end{equation}
Evaluated at thermal equilibrium, $P(T_{\rm eq})=p$ is the pressure of the fluid. The energy-momentum of the ideal fluid is obtained from the variation of the action with respect to the background metric
\begin{equation}
    T_{\rm f}^{\mu\nu}=\frac{2}{\sqrt{-g}}\frac{\delta S_{\rm f}}{\delta g_{\mu\nu}}=p P^{\mu\nu}+\varepsilon u^\mu u^\nu,
\end{equation}
where $P^{\mu\nu}=g^{\mu\nu}+u^\mu u^\nu$ is the projector on the directions transverse to the fluid velocity and we have used $\varepsilon=Ts-p=T\partial_T p-p$.

This can be extended beyond ideal hydrodynamics by adding derivative terms to the action. However, not all terms in the hydrodynamic equations can be captured by this approach, in particular not, dissipative terms. Generalizations that account for dissipation based on the Schwinger-Keldysh formalism and the Kubo-Martin-Schwinger condition in statistical field theory have been proposed in e.g.~\cite{Haehl:2015pja,Crossley:2015evo,Haehl:2015uoc,Glorioso:2017fpd,Haehl:2018lcu,Jain:2020hcu,Jain:2023obu}, but we will simply add dissipative terms by hand to the equations after deriving the non-dissipative part from the effective action.

To second order in derivatives of the scalar, the most general form of the action is
\begin{equation}
    S=\int d^d x \sqrt{-g} \left(\frac{Z_0}{2}(u^\mu\partial_\mu \phi)^2-\frac{Z_s}{2} P^{\mu\nu}\partial_\mu \phi\partial_\nu \phi -V\right).
\end{equation}
The coefficients in the kinetic terms of the scalar and the potential are both functions of the scalar and the temperature:
\begin{equation}
    Z_0=Z_0(T,\phi),\quad Z_s=Z_s(T,\phi),\quad V=V(T,\phi).
\end{equation}
In principle, we could add terms that have two derivatives of the temperature or the velocity, but those will not be contributing at leading order in the gradient expansion. Terms with a single derivative along the fluid velocity direction are forbidden by time reversal invariance. 

If we interpret $S$ as the effective action of the scalar in the presence of the fluid, the equation of motion of the scalar will be the Euler-Lagrange equation derived from the action:
\begin{equation}\label{eq:scalareq}
\partial_\mu\left(Z_0\, u^\mu u^\nu \partial_\nu\phi\right)
- \partial_\mu\left(Z_s\, P^{\mu\nu}\partial_\nu\phi\right)
- \frac{1}{2}\,\partial_\phi Z_0\,(u^\mu\partial_\mu\phi)^2
+ \frac{1}{2}\,\partial_\phi Z_s\,P^{\mu\nu}\partial_\mu\phi\,\partial_\nu\phi
+ \partial_\phi V + \nu_{\phi} = 0 .
\end{equation}
We have introduced in the equation an additional term, $\nu_{\phi}$, that does not originate in the action and that contains possible dissipative terms. The weak-coupling analysis suggests that, to lowest order in derivatives,
\begin{equation}
    \nu_{\phi}\approx \gamma u^\mu\partial_\mu \phi.
\end{equation}
With this term, the equation of motion for the scalar resembles the Josephson condition for a superfluid, with the derivative of the potential playing the role of the chemical potential. Note that this is not the most general dissipative term that we could add at this order; additional contributions proportional to $u^\mu\partial_\mu T$ and $\partial_\mu u^\mu$ could also have been introduced at this point, although the former can be traded for the latter using the leading order equations of motion~\cite{Gouteraux:2023uff}, so that only one of them is independent.

The energy-momentum tensor of the coupled scalar--fluid system is derived as before, from the variation with respect to the background metric
\begin{equation}\label{eq:Tmunu}
    T^{\mu\nu}=\frac{2}{\sqrt{-g}}\frac{\delta S}{\delta g_{\mu\nu}}+({\rm dissipative})=-V g^{\mu\nu}-T\partial_T V u^\mu u^\nu+T^{\mu\nu}_\phi+\pi^{\mu\nu},
\end{equation}
where we have added a term $\pi^{\mu\nu}$ containing possible dissipative terms. In this formulation, the scalar energy-momentum tensor corresponding to the kinetic terms is 
\begin{equation}
\begin{split}
    T^{\mu\nu}_\phi= &  Z_0 u^\mu u^\nu \dot\phi^2+Z_s\xi^\mu \xi^\nu-Z_s \xi^\mu u^\nu \dot\phi-Z_s \xi^\nu u^\mu \dot\phi\\
    &+\frac{1}{2}g^{\mu\nu}\left(Z_0 \dot\phi^2-Z_s\xi^2 \right)+\frac{1}{2}u^\mu u^\nu\left(T\partial_T Z_0 \dot\phi^2-T\partial_T Z_s\xi^2 \right),
\end{split}
\end{equation}
where we have introduced
\begin{equation}
    \dot\phi\equiv u^\mu\partial_\mu \phi,\quad \xi^\mu\equiv P^{\mu\nu}\partial_\nu \phi,\quad \xi^2=\xi^\mu \xi_\mu,
\end{equation}
in order to make the formulas more compact. It can be shown that
\begin{equation}
    u_\nu\partial_\mu T^{\mu\nu}_\phi =\Big(\partial_\mu(Z_s \xi^\mu)-\partial_\mu(Z_0 u^\mu) \Big)\dot\phi-T\partial_\mu\left[\left( \frac{\dot\phi^2}{2}\partial_T Z_0-\frac{\xi^2}{2}\partial_T Z_s \right)u^\mu\right]+\left(\frac{\dot\phi^2}{2}\partial_\phi Z_0-\frac{\xi^2}{2}\partial_\phi Z_s\right) \dot\phi
\end{equation}
Then, the conservation equation projected along the velocity is
\begin{equation}
\begin{split}
0=&   u_\nu \partial_\mu T^{\mu\nu}=T\partial_\mu\left[\left(\partial_T V +\frac{\xi^2}{2}\partial_T Z_s -\frac{\dot\phi^2}{2}\partial_T Z_0 \right)u^\mu\right]\\
&+\Big(\partial_\mu(Z_s \xi^\mu)-\partial_\mu(Z_0 u^\mu) -\partial_\phi V\Big)\dot\phi+\left(\frac{\dot\phi^2}{2}\partial_\phi Z_0-\frac{\xi^2}{2}\partial_\phi Z_s\right) \dot\phi+\partial_\mu \pi^{\mu\nu}u_\nu.
\end{split}
\end{equation}
The first line in the equation above allows for the identification of the entropy current as
\begin{equation}\label{eq:entropycurrent}
    s^\mu=\left(-\partial_T V +\frac{\dot\phi^2}{2}\partial_T Z_0 -\frac{\xi^2}{2}\partial_T Z_s\right)u^\mu .
\end{equation}
Using the equation of motion for the scalar, the divergence of the entropy current becomes
\begin{equation}
    T\partial_\mu s^\mu=\nu_{\phi} \dot\phi+\partial_\mu \pi^{\mu\nu}u_\nu\,.
\end{equation}

Imposing the Landau frame for the fluid component, $\pi^{\mu\nu}u_\nu=0$, we can rewrite $\partial_\mu \pi^{\mu\nu}u_\nu=-\pi^{\mu\nu}\partial_\mu u_\nu$, in which case the form of the dissipative terms that is consistent with the second law is, at first order in derivatives of the fluid velocity and the scalar field,\footnote{Note that this analysis discards adding the other contributions to $\nu_{\phi}$ mentioned above, proportional to  $u^\mu\partial_\mu T$ or $\partial_\mu u^\mu$.}
\begin{subequations}\label{eq:dissipativeFluid}
\begin{eqnarray}
    \nu_{\phi} &=&\gamma \dot\phi,\\
    \pi^{\mu\nu} &=& -\eta \sigma^{\mu\nu}-\zeta P^{\mu\nu}\partial_\alpha u^\alpha,
\end{eqnarray}
\end{subequations}
where $\eta$, $\zeta$ are the shear and bulk viscosities and the shear tensor is
\begin{equation}
    \sigma^{\mu\nu}=P^{\mu\alpha}P^{\nu\beta}\left(\partial_\alpha u_\beta+\partial_\beta u_\alpha-\frac{2}{d-1}P_{\alpha\beta}\partial_\lambda u^\lambda \right).
\end{equation}
With the identification of the dissipative terms we have done above, we arrive at an equation that determines the divergence of the entropy current consistent with the second law:
\begin{eqnarray}
    T\partial_\mu s^\mu=\gamma(u^\mu \partial_\mu \phi)^2+\frac{\eta}{2} \sigma^{\mu\nu}\sigma_{\mu\nu}+\zeta (\partial_\mu u^\mu)^2.
\end{eqnarray}
In the case where the kinetic terms are the Lorentz invariant terms of a free scalar, $Z_0=Z_s=1$, we recover the same formulas as for the weakly coupled case discussed at the beginning, with the addition of the viscosity terms.

\subsection{Thermodynamic equilibrium and linearized fluctuations}\label{sec:hydrofluc}

The scalar equation \eqref{eq:scalareq} implies that equilibrium states of constant temperature and scalar profile $\phi=\phi_{eq}(T)$ are extrema of the effective potential
\begin{equation}
    \partial_\phi V(\phi,T)\Big|_{\phi=\phi_{eq}}=0.
\end{equation}
From the energy-momentum tensor \eqref{eq:Tmunu} we can identify the pressure of the equilibrium configuration 
\begin{equation}
    p(T)=-V(\phi_{eq},T).
\end{equation}
The entropy density and energy density at equilibrium are obtained from the usual thermodynamic relations. The entropy density is
\begin{equation}
    s(T)=\frac{d}{dT}p(T)=-\partial_\phi V\Big|_{\phi=\phi_{eq}}\frac{d\phi_{eq}}{dT}-\partial_T V\Big|_{\phi=\phi_{eq}}=-\partial_T V\Big|_{\phi=\phi_{eq}},
\end{equation}
in agreement with the entropy current that we identified in \eqref{eq:entropycurrent}. The energy density is determined as
\begin{equation}
    \varepsilon(T)=T s-p=V-T\partial_T V\Big|_{\phi=\phi_{eq}},
\end{equation}
which also agrees with the form of the energy-momentum tensor given in \eqref{eq:Tmunu}.

Note that the temperature derivative is being taken along equilibrium configurations. We then have the constraint
\begin{equation}
    \frac{d}{dT} \partial_\phi V(\phi_{eq},T)=0.
    \label{eq: constraint V 1}
\end{equation}
This leads to
\begin{equation}
   \frac{d\phi_{eq}}{dT}=-\frac{\partial_T\partial_\phi V}{\partial_\phi^2 V}\Bigg|_{\phi=\phi_{eq}}.
   \label{eq: constraint V 2}
\end{equation}
We can use this expression to determine the specific heat
\begin{equation}\label{eq:cv}
    c_v=T\frac{ds}{dT}=-T\left( \partial_T^2V+\partial_T\partial_\phi V \frac{d\phi_{eq}}{dT}\right)\Bigg|_{\phi=\phi_{eq}}=\left(C+T\frac{A^2}{\partial_\phi^2V}\right)\Bigg|_{\phi=\phi_{eq}}.
\end{equation}
Here we have defined $C=-T\partial_T^2V$ and $A=-\partial_\phi\partial_T V$, where $C$ corresponds to the specific heat with the scalar frozen and $A$ accounts for the coupling between the scalar field and the fluid.

Let us now focus on the case where there is a first-order phase transition. As we vary the temperature, the number of extrema of the potential changes, from one at very high or low temperatures, to three at intermediate temperatures close to the value where the phase transition takes place. Each extremum corresponds to a different thermodynamic phase. When there is a single extremum, it is a minimum of the effective potential and the phase is stable. When there are three extrema, two are minima and one is a maximum. The two minima correspond to stable or metastable phases, while the maximum corresponds to the unstable phase, which we refer to as the spinodal branch. The transition from three extrema to one extremum occurs when a maximum and a minimum merge and go away, at the so-called spinodal points. 

If we move along equilibrium configurations from a metastable branch to the spinodal branch, we change from a minimum with $\partial_\phi^2 V>0$ to a maximum with $\partial^2_\phi V<0$, so at the spinodal points $\partial_\phi^2 V$ crosses zero. From \eqref{eq:cv}, one can see that the specific heat diverges $|c_v|\to \infty$ with the same sign as $\partial_\phi^2 V$. This implies that the spinodal branch is always thermodynamically unstable close enough to the spinodal points.

On the other hand, in the center of the spinodal branch the specific heat can be positive, which would appear to signal thermodynamic stability. It has been noted that one could treat the source for the scalar order parameter as a thermodynamic variable\footnote{We thank Blaise Goutéraux for pointing this out.} --- albeit one that is \emph{not} associated to a conserved charge \cite{Friess:2005zp,Gouteraux:2024adm}. 
Consider then the free energy as a function of the temperature and scalar source, formed by a Legendre transform of the potential $V(T,\phi)$, \textit{i.e.}  $F(T,J)=V(T,\phi)-\phi J$. Here $\phi(T,J)$ is defined implicitly as the solution of $\partial_\phi V(T,\phi)=J$, i.e.\ the value of the order parameter in equilibrium with source $J$ at temperature $T$. Taking the total differential of this relation,
\begin{equation}
    dJ = -A\,dT + \partial_\phi^2 V\,d\phi\,,
\end{equation}
and setting $dT=0$ or $dJ=0$ gives
\begin{equation}
    \left.\partial_J\phi\right|_T=\frac{1}{\partial_\phi^2 V}\,,\qquad
    \left.\partial_T\phi\right|_J=\frac{A}{\partial_\phi^2 V}\,.
\end{equation}
Using $J=\partial_\phi V$, the first derivatives of $F$ are simply
\begin{equation}
    \left.\partial_T F\right|_J=\left.\partial_T V\right|_\phi\,,\qquad
    \left.\partial_J F\right|_T=-\phi\,.
\end{equation}
One can then consider the two-by-two susceptibility matrix given by (minus) the Hessian of $F$,
\begin{equation}\label{eq:suceptibility}
    {\cal H}(T,J)\equiv -\begin{pmatrix}
    \partial_T^2 F & \partial_T\partial_J F \\
    \partial_J\partial_T F & \partial_J^2 F
    \end{pmatrix}
    =\begin{pmatrix}
    C/T+A^2/\partial_\phi^2 V & A/\partial_\phi^2 V \\
    A/\partial_\phi^2 V & 1/\partial_\phi^2 V
    \end{pmatrix},
\end{equation}
where derivatives of $F$ are taken at fixed $(T,J)$ and those of $V$ at fixed $(T,\phi)$. In analogy with the usual notion of local thermodynamic stability, we can require ${\cal H}$ defined in this way to be positive definite. Its determinant equals $C/(T\partial_\phi^2 V)$. Thus, assuming $C>0$, we have ``generalized'' thermodynamic stability as long as the potential is convex. On the spinodal branch, where $\partial_\phi^2 V<0$, ${\cal H}$ always has exactly one unstable eigenvalue, signaling instability.

To see this instability in the dynamics, we consider small fluctuations around one of the equilibrium states:
\begin{equation}
T\to T+\delta T(t,x) \ ,\quad \quad \phi=\phi_{eq}(T)+\delta \phi(t,x) \ , \quad \quad \ u^\mu=(1, \delta v^i(t,x)) \ .
\end{equation}
From the scalar equation \eqref{eq:scalareq} we obtain the following equation to linear order in the fluctuations
\begin{equation}
    Z_0 \partial_t^2 \delta \phi-Z_s \nabla^2 \delta \phi+\partial_\phi^2 V \delta\phi+\partial_\phi \partial_T V\delta T+\gamma \partial_t \delta\phi= 0.
\end{equation}
All coefficients here and below are evaluated at the equilibrium value of the scalar $\phi=\phi_{eq}(T)$.

The components of the energy momentum tensor \eqref{eq:Tmunu} expanded to linear order in the fluctuations are
\begin{equation}
\begin{split}
&T^{00}\approx \varepsilon-T\partial_\phi \partial_T V\,\delta\phi-T\partial_T^2 V\, \delta T,\\
&T^{0i}\approx (\varepsilon+p)\, \delta v^i,\\
&T^{ij}\approx (p-\partial_T V\,\delta T)\delta^{ij}-\eta \left(\partial^i \delta v^j+\partial^j \delta v^i-\frac{2}{d-1}\delta^{ij}\partial_k \delta v^k\right)-\zeta\delta^{ij} \partial_k \delta v^k.
\end{split}
\end{equation}
Introducing the expressions above in the conservation equations for energy and momentum, we find
\begin{eqnarray}\label{eq:conserveqs}
 0&=& T\partial_\phi \partial_T V\,\partial_t \delta\phi+T\partial_T^2 V\, \partial_t \delta T-(\varepsilon+p)\partial_i \delta v^i,\\
 0&=& (\varepsilon+p)\partial_t\delta  v^i-\partial_T V\,\partial^i \delta T-\eta \left(\partial^2 \delta v^i+\frac{d-3}{d-1}\partial^i\partial_k \delta v^k \right)-\zeta \partial^i \partial_k \delta v^k.
\end{eqnarray}
We expand the fluctuations in plane waves
\begin{equation}
    \delta X(t,x)=\int \frac{d\omega d^{d-1}q}{(2\pi)^d}e^{-i\omega t +i q\cdot x} \delta X(\omega,q)
\end{equation}
and solve the resulting algebraic equations for the Fourier transform $\delta X(\omega,q)$.

We first determine the longitudinal component of the velocity $\partial\cdot \delta v=\partial_i\delta  v^i$ from the momentum conservation equation 
\begin{equation}
   \partial\cdot \delta v=\frac{1}{T}\frac{q^2}{-i \omega+\Gamma q^2}\delta T,\quad \Gamma=\frac{1}{T}\left( \frac{\zeta}{s}+2\frac{d-2}{d-1}\frac{\eta}{s}\right).
\end{equation}
After introducing this into the energy conservation equation, we solve for the temperature fluctuation 
\begin{equation}
    \delta T=\frac{i\omega T A}{-i\omega C+\frac{s q^2}{-i\omega+\Gamma q^2}}\delta\phi.
\end{equation}
Finally, the scalar equation gives a condition that will determine the dispersion relation of the modes
\begin{equation}
   0= -\omega^2 Z_0+q^2 Z_s+\partial_\phi^2 V-i\omega \gamma-\frac{i\omega T A^2}{-i\omega C+\frac{s q^2}{-i\omega+\Gamma q^2}}
\end{equation}
Using \eqref{eq:cv} to eliminate $A$, this can be massaged to give
\begin{equation}\label{eq:dispersion}
\begin{split}
  0=& \ \omega^4+i \omega^3 \left(\frac{\gamma}{Z_0}+q^2 \Gamma \right)-\omega^2\left(\frac{\partial_\phi^2V}{Z_0} \frac{c_v}{C}+q^2\left(\frac{Z_s}{Z_0}+ \frac{s}{C}+\frac{\gamma}{Z_0} \Gamma\right) \right)\\
  &-i \omega q^2\left( \Gamma \frac{\partial_\phi^2 V}{Z_0}\frac{c_v}{C}+\frac{\gamma}{Z_0}\frac{s}{C}+q^2 \frac{Z_s}{Z_0} \Gamma \right)+q^2\frac{s}{ C}\left( \frac{\partial_\phi^2 V}{Z_0}+q^2 \frac{Z_s}{Z_0}\right) \ .
  \end{split}
\end{equation}
This dispersion relation depends on the following combinations of thermodynamic quantities, terms from the effective action and transport coefficients:
\begin{equation}
    \frac{\gamma}{Z_0},\ \frac{Z_s}{Z_0},\ \frac{\partial_\phi^2 V}{Z_0},\ \frac{c_v}{C},\ \frac{s}{c_v},\ \Gamma.
    \label{eq: EFT parameters}
\end{equation}
We study next how stability and spatial correlations are affected as we move between different phases.

\subsection{Dynamical stability and correlation length}\label{sec:spinodal}

Two interesting regimes are the low-momentum regime $q/T\ll 1$, which includes hydrodynamic fluctuations, and the static $\omega=0$ limit, which encodes spatial correlations.

Let us start with the low momentum regime. When the spatial momentum vanishes $q= 0$ \eqref{eq:dispersion} becomes
\begin{equation}
    0\approx \omega^2\left(\omega^2+i\omega\frac{\gamma}{Z_0}-\frac{\partial_\phi^2V}{Z_0} \frac{c_v}{C}\right).
\end{equation}
The two $\omega=0$ roots correspond to gapless hydrodynamic modes. There are two other roots at non-zero complex frequencies, corresponding to gapped modes (relative to zero frequency)  with dispersion relations
\begin{equation}\label{eq:dispersionGapMode}
    \omega_\pm(q)=-i\frac{\gamma}{2Z_0}\pm \Delta +\alpha_{\pm,2} q^2+O(q^4),
\end{equation}
where
\begin{equation}
    \Delta=\sqrt{\frac{\partial_\phi^2V}{Z_0} \frac{c_v}{C}-\left(\frac{\gamma}{2Z_0}\right)^2} ,
\end{equation}
and
\begin{equation}
     \alpha_{\pm,2}=\frac{\frac{\partial_\phi^2V}{Z_0} \frac{c_v}{C}\left(\frac{Z_s}{Z_0}+\frac{s}{c_v}\left(\frac{c_v}{C}-1\right)\right)-i\frac{Z_s }{Z_0}\frac{\gamma}{Z_0}\left(-i\frac{\gamma}{2Z_0} \pm\Delta\right)  }{2\frac{\partial_\phi^2V}{Z_0} \frac{c_v}{C}\left(-i\frac{\gamma}{Z_0} \pm\Delta\right)-\frac{\gamma^2}{Z_0^2}\left(-i\frac{\gamma}{2Z_0} \pm\Delta\right)} \ .
\label{eq:alpha2}
\end{equation}
In order to determine the dispersion relation of the hydrodynamic modes we expand \eqref{eq:dispersion} for $\omega\sim q\ll T$. To leading order,
\begin{equation}
     0= -\omega^2\frac{\partial_\phi^2V}{Z_0} \frac{c_v}{C} +q^2\frac{s}{ C}\frac{\partial_\phi^2 V}{Z_0}+O(\omega q^2,\omega^3).
\end{equation}
The solutions are the sound modes, with dispersion relation 
\begin{equation}\label{eq:dispersionSound}
    \omega_s(q)=\pm \sqrt{\frac{s}{c_v}}|q|+O(q^2).
\end{equation}
In the stable branches $\partial_\phi^2 V>0$ and $c_v>0$, so all the modes have dispersion relations with a negative imaginary part for the frequency, and these states are dynamically stable, as the fluctuations decay with time. In the spinodal branch $\partial_\phi^2 V<0$. If the branch is thermodynamically unstable, $c_v<0$, as it happens close to the spinodal points, then the gapped modes \eqref{eq:dispersionGapMode} are still decaying, but one of the sound modes \eqref{eq:dispersionSound} has a positive imaginary part of the frequency and there is a spinodal instability. If at some point along the spinodal branch one recovers thermodynamic stability, $c_v>0$, then the sound-mode (spinodal) instability
disappears, but one of the gapped modes (with the plus sign in the square root) becomes exponentially growing, and the state is still dynamically unstable. The thermodynamic stability is thus linked to the dynamical stability of the scalar at these points.

The point where the nature of the dynamical instability of the spinodal branch switches is where the specific heat vanishes, $c_v=0$. Using \eqref{eq:cv}, if we assume $-T\partial_T^2 V>0$ in all configurations, then  $c_v$ can become positive in the spinodal branch, where $\partial_\phi^2 V<0$, at the values satisfying the bound
\begin{equation}
    |\partial_\phi^2 V|\geq \frac{(\partial_T\partial_\phi V)^2}{|\partial_T^2 V|}.
\end{equation}
At the spinodal points it is necessary to find the dispersion relation to the next order in the momentum expansion, which is straightforward to do from \eqref{eq:dispersion}:
\begin{equation}\label{eq:dispersionSound2}
    \omega_s(q)=\pm \sqrt{\frac{s}{c_v}}|q|-\frac{i}{2}\left(\Gamma+\frac{\gamma s}{c_v \partial^2_\phi V}\frac{c_v-C}{c_v}\right)q^2+O(|q|^3)
\end{equation}
When $|c_v|\to \infty$, $\partial_\phi^2 V\to 0$, and taking into account \eqref{eq:cv}, the dispersion relation of the sound modes becomes
\begin{equation}
\omega_s(q)=-\frac{i}{2}\left(\Gamma+\frac{\gamma s}{T A^2}\right)q^2+O(|q|^3).
\end{equation}
Therefore the spinodal points are marginally stable with respect to linearized fluctuations.

Moving on to the zero frequency dispersion relation, from \eqref{eq:dispersion} we obtain the condition
\begin{eqnarray}
    0=q^2\frac{s}{C}\left( \frac{\partial_\phi^2 V}{Z_0}+q^2 \frac{Z_s}{Z_0}\right).
\end{eqnarray}
The two modes with $q^2=0$ correspond to the hydrodynamic sound modes. There are two additional modes that exist for spatial momentum
\begin{equation}\label{eq:momentumdisp}
    q_{\text{max}}^2=-\frac{\partial_\phi^2 V}{Z_s} \ .
\end{equation}
In the stable branches, where $\partial_\phi^2 V>0$, these are purely imaginary and describe an exponential decay of fluctuations in the spatial directions
\begin{equation}
    \delta X\sim e^{-|x|/\xi}.
\end{equation}
The  correlation length that determines the suppression is
\begin{equation}
    \xi=\left(\frac{Z_s}{\partial_\phi^2 V}\right)^{1/2}.
\end{equation}
We see that when one approaches the spinodal points from a metastable branch, $\partial_\phi^2 V\to 0$, and the correlation length diverges. This behavior is analogous to the divergence of the correlation length at critical points in second-order phase transitions. 

In the spinodal branch $\partial_\phi^2 V<0$, and the modes satisfying \eqref{eq:momentumdisp} have real spatial momentum. This corresponds to the highest value of momentum for which there is a spinodal instability. Then, when one approaches the spinodal points from the unstable branch, the range of momenta where spinodal instability is present decreases, until it disappears at the spinodal points. Note that setting $\omega=0$ in the hydrodynamic dispersion relation \eqref{eq:dispersionSound2} also gives a prediction for the range of momenta where there is a spinodal instability, which is however different. Thus, the endpoint of the spinodal instability in momentum must be outside the hydrodynamic regime.

\subsection{The probe approximation for the scalar}\label{sec:probe}

In the effective description we are using, we distinguish between a fluid component and the scalar field. If the number of degrees of freedom of the fluid is very large, for instance when $g_{\rm eff}\gg 1$ in \eqref{eq:weakpressure}, we might be able to neglect the contribution of the scalar field to the pressure and other thermodynamic quantities and treat it as a probe. In terms of the effective potential, there will be a hierarchy between the potential and its temperature derivatives relative to the scalar field derivatives
\begin{equation}
    |V|,|\partial_T V|,|\partial_T^2 V| \gg |\partial_\phi V|,|\partial_\phi^2 V|,|\partial_T\partial_\phi V|.
\end{equation}
It follows that the specific heat \eqref{eq:cv} in the probe approximation does not depend on the scalar field derivatives
\begin{equation}
    c_v\approx -T\partial_T^2 V = C,
\end{equation}
and, for a normal fluid, it would always be positive $c_v>0$. Thus, in this approximation there are no changes in the thermodynamic stability, in particular the spinodal branch is thermodynamically stable. Clearly, the probe approximation must break around the spinodal points, as the term with scalar field derivatives in \eqref{eq:cv} would not be negligible when $\partial_\phi^2V\to 0$. Nevertheless, the region in the phase diagram where it breaks down could be very small.

Physically, the probe approximation treats the plasma as a heat bath whose temperature is unaffected by the scalar. Near the spinodal points, however, the equilibrium value of the scalar becomes infinitely susceptible to temperature changes, $d\phi_{eq}/dT=-\partial_T\partial_\phi V/\partial_\phi^2V$ (see \eqref{eq: constraint V 2}). However weak the coupling $\partial_T\partial_\phi V$, the heat absorbed or released by the condensate then dominates the specific heat, and it feeds back on the temperature of the plasma.

Concerning the dynamical stability of the equilibrium states, the linearized equations for the fluctuations \eqref{eq:scalareq} and \eqref{eq:conserveqs} would be
\begin{eqnarray}
 0&=& Z_0 \partial_t^2 \delta \phi-Z_s \nabla^2 \delta \phi+\partial_\phi^2 V \delta\phi+\partial_\phi \partial_T V\delta T+\gamma \partial_t \delta\phi,\\
 0&\approx & T\partial_T^2 V\, \partial_t \delta T-(\varepsilon+p)\partial_i \delta v^i,\\
 0&=& (\varepsilon+p)\partial_t \delta v^i-\partial_T V\,\partial^i \delta T-\eta \left(\partial^2  \delta v^i+\frac{d-3}{d-1}\partial^i\partial_k \delta v^k \right)-\zeta \partial^i \partial_k\delta  v^k.
\end{eqnarray}
The temperature and velocity fluctuations decouple from the scalar and their equations do not depend on scalar field derivatives of the effective potential. Since the specific heat remains positive, there would be no spinodal instability associated with the sound mode. On the other hand, the fluctuations of the scalar can be studied setting the temperature and velocity fluctuations to zero and lead to the condition for the frequency and spatial momentum of the mode which is
\begin{equation}
 0=Z_0\omega^2-Z_s q^2-\partial_\phi^2 V+i\gamma\omega.
\end{equation}
In this case, the dispersion relation is
\begin{equation}
    \omega=-i\frac{\gamma}{2Z_0}\pm\sqrt{\frac{\partial_\phi^2V}{Z_0}+\frac{Z_s}{Z_0}q^2-\left(\frac{\gamma}{2Z_0}\right)^2}.
\end{equation}
At $q=0$ this coincides with \eqref{eq:dispersionGapMode} when $c_v=C$. On the spinodal branch, where $\partial_\phi^2V<0$, the $+$ root is purely imaginary with $\mathrm{Im}\,\omega>0$ for $0\le q<q_{\text{max}}$. Here $q_{\text{max}}$ is the same as in the backreacted case, \eqref{eq:momentumdisp}. The probe limit thus reproduces the gapped order-parameter instability that the full theory exhibits where $c_v>0$, together with the width of the unstable band.

It misses two effects of the coupling to hydrodynamics. First, near the spinodal points, where $c_v<0$, the full theory transfers the instability from the gapped mode to the sound channel; this switch is absent in the probe limit. Second, in the probe limit the growth rate is maximal at $q=0$, whereas with backreaction the maximum can shift to $q\neq0$ even while the gapped mode carries the instability (see e.g. Fig.~\ref{fig:QNM-vs-EFT-A-4} below).

\section{Holographic dual to a first-order phase transition}\label{sec:holog}

In order to test the validity of the effective theory we will do a microscopic calculation of the modes corresponding to the fluctuations around thermal equilibrium of a system undergoing a first-order phase transition. We will use the  holographic model developed in \cite{Bea:2018whf} to study the thermodynamic and dynamical stability of the spinodal branch. The gravitational theory consists of five-dimensional Einstein gravity coupled to a real scalar field, dual to a $d=4$ field theory,
\begin{equation}\label{eq:gravityact}
S\,=\,\frac{1}{4\pi G_5}\int \dd^5x\sqrt{-g}\left(\frac14 R-\frac12 \partial_M\psi\partial^M\psi
-\mathcal{V}(\psi)\right)\,, 
\end{equation}
where $G_5$ is the five-dimensional Newton constant. The potential takes the form
\begin{equation}
    \label{eq:V}
\mathcal{V}(\psi)= -\frac{4}{3}W(\psi)^2 + \frac{1}{2}\left(\frac{\partial W(\psi)}{\partial \psi}\right)^2\,,
\end{equation}
where
\begin{equation}\label{eq:superpot}
    LW(\psi) = -\frac{3}{2} -\frac{\psi^2}{2} -  \frac{\psi^4}{4\psi_M^2} +  \frac{\psi^6}{\psi_Q}\,.
\end{equation}
The function $W$ is usually known as the superpotential, although there is no supersymmetry involved here. At $\psi=0$ there is a maximum of both the potential and the superpotential
\begin{equation}\label{eq:expansion_potential}
    L^2\mathcal{V}(\psi) \approx -3 -\frac{3}{2}\psi^2\,.
\end{equation}
The $\psi=0$ solution is $AdS_5$ with radius $L$, which is dual to a four-dimensional strongly coupled CFT. We see that the scalar has a mass $m^2L^2=-3$ around this point. Following the usual holographic dictionary, this means that the field is dual to a scalar operator of conformal dimension $\Delta=3$.

The gravity dual of thermal equilibrium states are black-brane static solutions to the equations of motion derived from \eqref{eq:gravityact}, which have to be found numerically. For that, we introduce the following metric ansatz,
\begin{equation}\label{eq: metric ansatz schwarzschild}
    ds^2 = \frac{L^2}{r^2}\left(-f(r)g(r)dt^2 + \frac{dr^2}{f(r)}+g(r)d\bm{x}^2\right), \quad \psi = \psi(r),
\end{equation}
with $r$ being the holographic radial direction. The AdS$_5$ boundary is located at $r=0$ and the horizon at $r = r_H$, where $f(r_H)=0$. 
Near the boundary, the different functions behave as
\begin{equation}\label{eq:boundexp}
    \begin{aligned}
        f & = 1 + f_4 r^4 + \mathcal{O}\left(r^6\right),\\
        g  & = 1 - \frac13 \psi_s^2 r^2 + \left(\frac{\psi_s^4}{18}-\frac{1}{2}\psi_s\psi_2\right)r^4+\mathcal{O}\left(r^6\right),\\
        \psi & =  \psi_s r + \psi_2 r^3 + \mathcal{O}\left(r^5\right).
    \end{aligned}
\end{equation}
The coupling $\psi_s=\Lambda$ defines different theories, so we want to set it to some constant value across all holographic solutions.

Near the horizon, all functions are regular and can be Taylor expanded around $r=r_H$. For numerical convenience, we set the radius of the horizon to unity, $r_H = 1$, by exploiting the symmetry 
\begin{equation}
    (t,r,\bm x)\rightarrow \lambda (t,r,\bm x),
\end{equation}
and redefine our scalar field $\psi = r\tilde\psi(r)$, so that it does not vanish at the boundary, $\tilde\psi(0) = \psi_s$.

As boundary conditions, we impose for the metric regularity for all functions at the horizon, with $f(1)=0$, and that the asymptotic AdS spacetime is properly normalized, i.e. $f(0)=g(0)=1$. The remaining condition on the scalar field can be imposed through either the value of $\psi_s$ or its value at the horizon, $\psi_H$. The latter is more convenient in exploring phase diagrams as we know that its value is contained in the range, $0\leq \psi_H \leq \psi_\star$, where $\psi_{\star}$ corresponds to the minimum of the potential, $\mathcal{V}'(\psi_\star) = 0$, in \eqref{eq:V}.

\subsection{Phase diagram and thermodynamics}

In order to determine the phase diagram we have to extract the values of thermodynamic quantities for the solutions, in particular the free energy and the temperature. For a fixed temperature, the state with lowest free energy will be thermodynamically preferred and other possible competing states would be unstable or metastable. The first-order phase transition will happen when the free energies of two states cross as the temperature is varied.

The coefficients in the boundary expansions \eqref{eq:boundexp} determine the expectation value of the energy-momentum and scalar operators, the energy density $\varepsilon$ and the pressure $p$. We have collected the formulas in appendix~\ref{app:holoren}. 
The free energy density equals minus the pressure in this setup $f=-p$. On the other hand, from the expansion of the solutions at the horizon, the entropy density, $s$, and temperature, $T$, of the black brane are extracted from the area of the horizon, $A_H$, and the surface gravity as\footnote{ Factors of $r_H$ cancel out when we compute dimensionless quantities, thus these can be computed directly from the numerical solution with $r_H=1$.}
\begin{equation}\label{eq:thermo_horizon_exp}
 \frac{4\pi G_5}{L^3} s \equiv \frac{4\pi G_5}{L^3}\frac{A_H}{4G_5} = \frac{\pi g(r_H)^{3/2}}{r_H^3},\quad T = \frac{|f'(r_H)|\sqrt{g(r_H)}}{4\pi}
\end{equation}
We have divided the entropy density by the factor $L^3/(4\pi G_5)$ that is proportional to the number of degrees of freedom of the dual field theory. One can check that the entropy density obtained from the horizon satisfies the expected thermodynamic identities
\begin{equation}
    \varepsilon+p=Ts,\quad s=\frac{\partial p}{\partial T}.
\end{equation}

From these quantities and the value of the scalar field at the horizon, $\psi_H$, one can obtain the specific heat ($c_v$), and the shear ($\eta$) \cite{Kovtun:2004de} and bulk ($\zeta_s$) \cite{Eling:2011ms} viscosities 
\begin{equation}
c_v=T\frac{\partial s}{\partial T},\quad     \eta=\frac{s}{4\pi},\quad \frac{\zeta_s}{\eta}=4\left(s\frac{\partial \psi_H}{\partial s}\right)^2\,.
\end{equation}
In general $\zeta_s$ is {\em different} from the bulk viscosity $\zeta$ in the constitutive relations \eqref{eq:dissipativeFluid}, but it determines the sound attenuation constant (in $d=4$ dimensions)
\begin{equation}\label{eq:GammaS}
    \Gamma_s=\frac{1}{Ts}\left(\zeta_s+\frac{4}{3}\eta\right).
\end{equation}
The speed of sound ($c_s$) squared in turn is given by
\begin{equation}
\label{eq:cs2fromthermo}
    c_s^2=\frac{\partial p}{\partial \varepsilon}=\frac{s}{c_v}.
\end{equation}
In section~\ref{sec:extractingEFTcoeffs}, we will compare these formulas with the results from the dispersion relations of hydrodynamic sound modes.

For different values of $\psi_M$ and $\psi_Q$ in the superpotential \eqref{eq:superpot}, we obtain different phase diagrams. The different families of phase diagrams that are obtained for different fixed $\psi_Q$ are qualitatively analogous to each other, so we will keep $\psi_Q = 10$ fixed. Then, for $\psi_M \lesssim 1.1$ we obtain first-order phase transitions. Greater values give crossovers, and in between there is one value, $\psi_M\approx1.1$, which results in a second-order phase transition.

FOPTs exhibit various competing states at the same temperature, rendering some of these states thermodynamically unstable. These instabilities can be either global or local. The former correspond to metastable states not minimizing the appropriate thermodynamic potential for the chosen ensemble. Local thermodynamic instabilities are those associated to states with the wrong sign in some susceptibility, in our case the specific heat. 

\begin{figure}[ht!]
    \centering
    \includegraphics[width=0.49\linewidth]{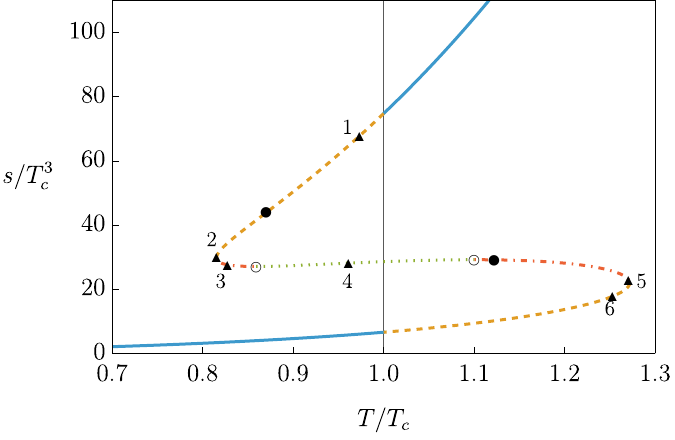}\hfill
    \includegraphics[width=0.47\linewidth]{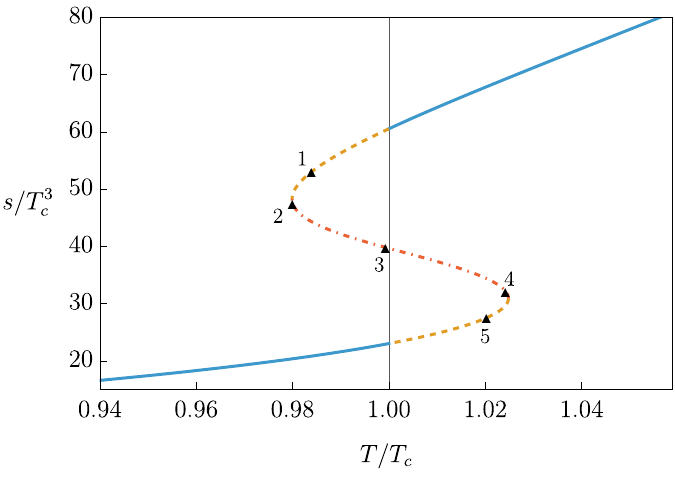}\\
    \caption{Entropy density as a function of temperature for two FOPTs in which $c_v$  changes sign along the spinodal branch (left) or remains negative (right). We refer to these theories as A (left) and B (right). The critical temperature is $T_c = 0.430\Lambda$ and $T_c = 0.396\Lambda$ for theories A and B respectively.} 
    \label{fig:s-of-T}
\end{figure}

Among the first-order phase transitions, there are two classes. Those for which the specific heat does not switch sign along the spinodal branch, when $\psi_M \geq 0.839$, and those for which it does change sign, $\psi_M < 0.839$. In this work, we select one example from each class: one with $\psi_M=0.8$, which we call \emph{theory A}, and one with $\psi_M=1$, which we call \emph{theory B}.
Their phase diagrams are shown in Fig.~\ref{fig:s-of-T}, where we plot the entropy as a function of temperature normalized by the critical temperature. The critical temperature is obtained by finding the pair of states that have, simultaneously, same temperature and pressure, which leads to $T_c = 0.430\Lambda$ for theory A and $T_c = 0.396\Lambda$ for theory B.

In Fig.~\ref{fig:s-of-T}, we display states that are globally stable with a solid blue line, those that are metastable in dashed yellow, those with negative specific heat, $c_v <0$, in dot-dashed red and those with $c_v > 0$ in the spinodal branch in dotted green.

\begin{figure}[ht!]
    \centering
    \includegraphics[width=0.49\linewidth]{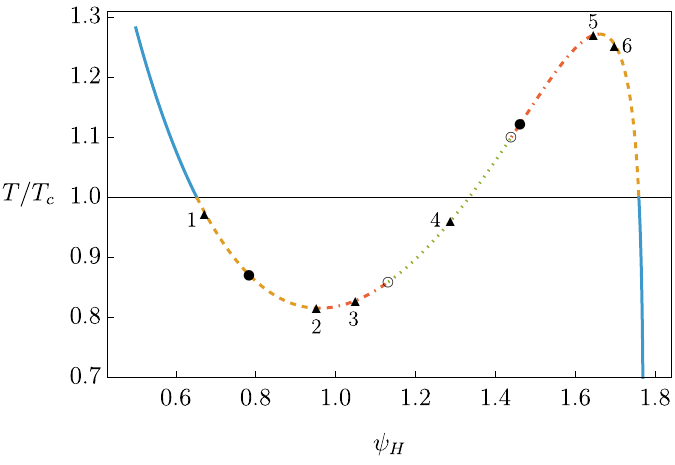}\hfill
    \includegraphics[width=0.49\linewidth]{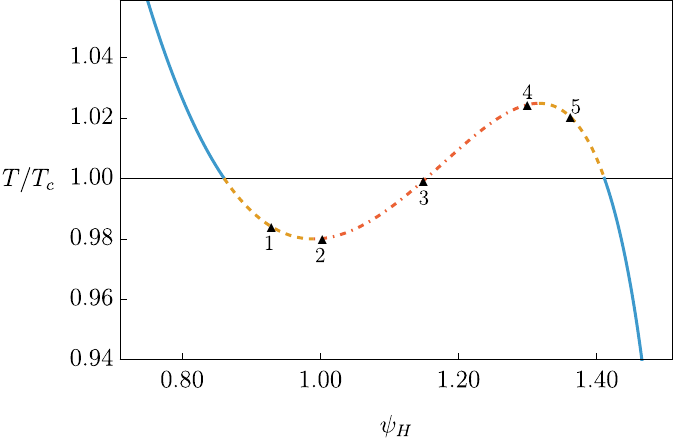}\\
    \caption{Value of the temperature as a function of the scalar field at the horizon for theory A (left) and theory B (right). On the left plot, $c_v$ changes sign along the spinodal branch, whereas on the right one it remains negative.}
    \label{fig:psiH-of-T}
\end{figure}

It is useful to have a parameter for the family of states that is always single-valued. For our holographic geometries, such a parameter is given by the horizon value of the scalar field $\psi_H$. In Fig.~\ref{fig:psiH-of-T}, temperature is plotted as a function of $\psi_H$, and the resulting curve is seen to be single-valued for both of the theories.

\subsection{Varying the backreaction}

As was explained in the previous section, for some values of the parameters, the spinodal branch of the model of \cite{Bea:2018whf} contains a region with positive specific heat, $c_v > 0$, where the speed of sound is real and the instability cannot be captured by hydrodynamic modes. In the probe approximation, where $c_v > 0$ away from the spinodal points, the effective theory predicts that the instability is carried by a non-hydrodynamic mode of the scalar field, which is unstable already at zero momentum. To understand how this picture changes as the backreaction is switched on, we interpolate between the probe limit and the fully backreacted theory.

To this end, we follow \cite{Mateos:2026moc} and introduce a parameter $\chi$ that controls the strength with which the scalar backreacts on the geometry. In the model of \cite{Bea:2018whf}, this amounts to placing $\chi$ in front of the matter sector of the action \eqref{eq:gravityact},
\begin{equation}\label{eq:gravityact_backreaction}
    \tilde{S} \,=\, \frac{1}{4\pi G_5} \int d^5x \sqrt{-g} \left(\frac{1}{4}R - \frac{1}{2}\Lambda_C - \frac{\chi}{2}\partial_M \psi \partial^M\psi - \chi\,\tilde{\mathcal{V}}(\psi)\right),
\end{equation}
where we have split the original potential as $\mathcal{V}(\psi) = \Lambda_C/2 + \tilde{\mathcal{V}}(\psi)$, with $\Lambda_C = -6/L^2$, so that $\chi$ does not multiply the cosmological constant. The resulting equations of motion read
\begin{equation*}
    G_{\mu\nu} + \Lambda_C g_{\mu\nu} \,=\, 2\chi\, T_{\mu\nu}\,, \qquad
    \Box_g\psi \,=\, \tilde{\mathcal{V}}'(\psi)\,,
\end{equation*}
with
\begin{equation*}
    T_{\mu\nu} \,=\, \partial_\mu\psi\,\partial_\nu\psi - g_{\mu\nu}\left(\frac{1}{2}\partial_\alpha\psi\,\partial^\alpha\psi + \tilde{\mathcal{V}}(\psi)\right)
\end{equation*}
the stress tensor of the scalar field. The parameter $\chi$ therefore controls how strongly the scalar sources the geometry, while leaving the scalar equation itself unchanged. In particular, the mass of the scalar and hence the dimension $\Delta = 3$ of the dual operator are independent of $\chi$. Setting $\chi = 1$ recovers the original model, while the limit $\chi \to 0$ corresponds to a probe scalar propagating on a fixed AdS-Schwarzschild background.

This parametrization is motivated by top-down constructions with $N_f$ flavor D-branes embedded in the background generated by $N_c$ color branes \cite{Karch:2002sh}. There, the gravitational action scales as $N_c^2$ while the brane action scales as $N_f N_c$. The backreaction of the branes on the geometry is therefore controlled by $\chi \sim N_f/N_c$, and the probe limit corresponds to $\chi \to 0$.

We use the same metric ansatz \eqref{eq: metric ansatz schwarzschild} as before. The near-boundary expansions \eqref{eq:boundexp} now acquire $\chi$-dependent terms:
\begin{equation}\label{eq:boundexp_backreaction}
    \begin{aligned}
        f & = 1 + f_4 r^4 + \mathcal{O}\left(r^6\right),\\
        g & = 1 - \frac13 \chi \psi_s^2 r^2 + \left(\frac{\chi(2\chi-1)}{18}\psi_s^4 - \frac{\chi}{2}\psi_s\psi_2 - \frac{\chi(\chi-1)}{3}\psi_s^4\log r\right)r^4 + \mathcal{O}\left(r^6\right),\\
        \psi & = \psi_s r + \left(\psi_2 + \frac{2}{3}(\chi-1)\psi_s^3\log r\right) r^3 + \mathcal{O}\left(r^5\right).
    \end{aligned}
\end{equation}
Logarithmic terms appear for $\chi \neq 1$, and setting $\chi = 1$ recovers \eqref{eq:boundexp}. The temperature and the entropy density are read off from horizon data, so their expressions \eqref{eq:thermo_horizon_exp} remain unchanged.

\begin{figure}[ht!]
    \centering
    \includegraphics[width=0.49\linewidth]{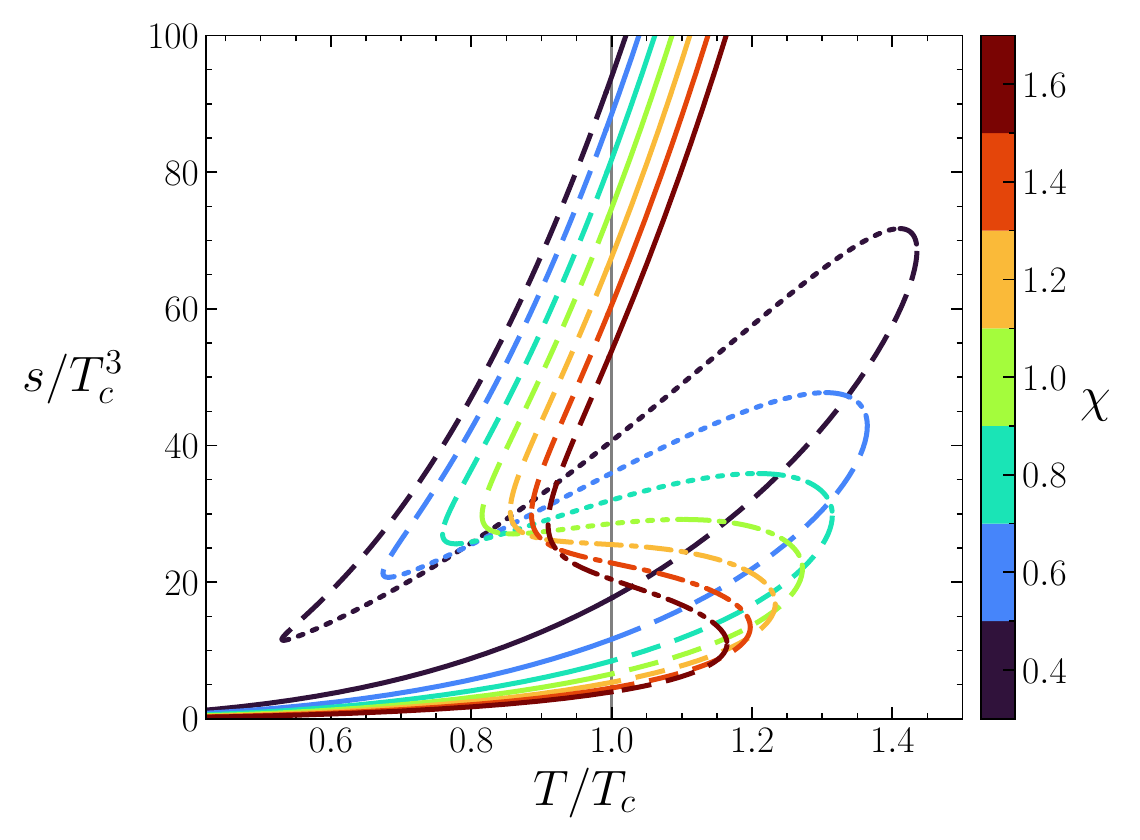}
    \includegraphics[width=0.49\linewidth]{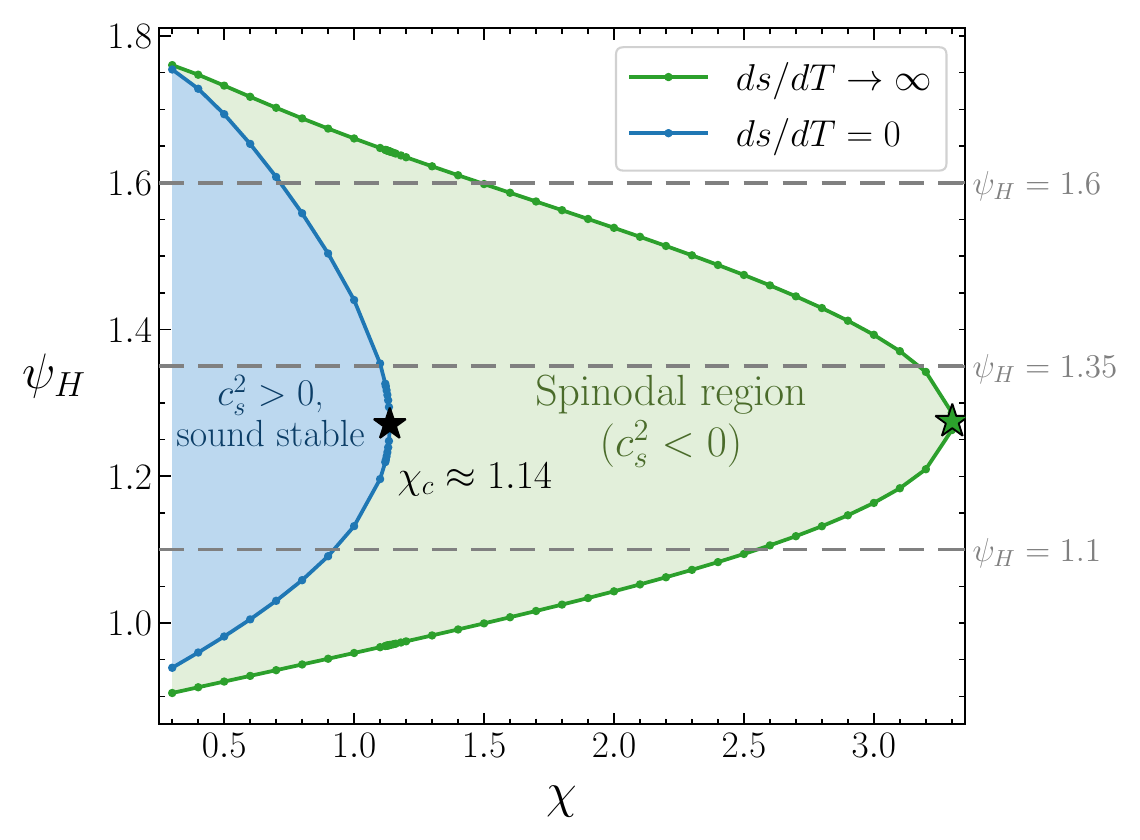}
    \caption{\textbf{Left:} entropy density as a function of temperature, normalized by the critical temperature $T_c$, for the holographic model with $\psi_M = 0.8$, $\psi_Q = 10$ and several values of the backreaction parameter $\chi$ (color-coded). Solid lines denote globally stable states, dashed lines metastable states, dot-dashed lines unstable states with $c_s^2 < 0$, and dotted lines unstable states with $c_s^2 > 0$. \textbf{Right:} value of the scalar field at the horizon, $\psi_H$, as a function of $\chi$. The green and blue curves mark $ds/dT \to \infty$ and $ds/dT = 0$, respectively. They bound the green region of states with $c_s^2 < 0$ and the blue region of unstable states with $c_s^2 > 0$. The black star indicates the value of $\chi$ at which the blue region closes, and the green star the value at which the first-order phase transition disappears. Dashed gray lines show the slices along which we study the quasinormal modes below.}
    \label{fig:SofT_backreaction}
\end{figure}

The left panel of Fig.~\ref{fig:SofT_backreaction} shows the entropy density as a function of the temperature for the theory with $\psi_M = 0.8$ and $\psi_Q = 10$ and several values of $\chi$. As we approach the probe limit $\chi \to 0$, the states with negative specific heat ($c_v < 0$) are squeezed towards the spinodal points, where the probe approximation is expected to break down. Meanwhile, the intermediate region of unstable states with positive specific heat ($c_v > 0$) becomes increasingly large. Conversely, this region shrinks as $\chi$ increases and disappears at the value of $\chi$ marked by the black star in the right panel. For large enough $\chi$, the first-order phase transition weakens until it becomes second-order at a critical point (green star). Beyond this value it turns into a smooth crossover. In field-theory terms, decreasing $\chi$ increases the number of spectator degrees of freedom relative to those that participate in the transition. We refer to appendix~A of \cite{Mateos:2026moc} for a detailed discussion of this interpretation.

A way to understand why the transition disappears when $\chi$ becomes large enough is to rescale the scalar field in \eqref{eq:gravityact_backreaction} to make it canonically normalized, $\psi\to \psi/\sqrt{\chi}$. This changes the potential term to $\chi \tilde {\cal V}(\psi/\sqrt{\chi})$. Then, when $\chi$ becomes large, the potential can be power expanded around $\psi/\sqrt{\chi}\approx 0$, in such a way that only the constant and quadratic contributions of the potential remain, while interaction terms of the scalar are suppressed by inverse powers of $\chi$. Thus, the action approaches that of a free scalar coupled to gravity with a cosmological constant, that exhibits no phase transitions.\footnote{We thank Elias Kiritsis for pointing this out.}

\section{Dynamical (in-)stability and comparison with the effective description}\label{sec:qnms}

In what follows we will study the linear, dynamical (in)stability of the states for the two holographic theories in Fig.~\ref{fig:s-of-T} and compare the results with the effective description presented in section~\ref{sec:scalar}. We will see that the effective description correctly predicts the various instabilities and gives a clear relation between linear dynamical stability and thermodynamic stability for these kinds of models.

\subsection{Quasinormal modes}

To determine the linear stability of the states, we compute the quasinormal mode (QNM) spectrum of the dual equilibrium black-brane configurations.

We follow standard techniques (see \cite{Jansen:2017oag} for an extensive discussion), perturbing the equilibrium geometries and recasting the linearized bulk equations of motion as a generalized eigenvalue problem. To do so, we start by perturbing the equilibrium configuration $\{\bar g_{MN},\bar\psi \}$ as
\begin{equation}
    g_{MN}(t,r,x) = \bar g_{MN}(r)+e^{-i(\omega t - q x)} h_{MN}(r),\quad \psi(t,r,x) = \bar\psi(r) + e^{-i(\omega t - q x)}\delta\psi(r),
\end{equation}
where we have perturbed our geometry with a specific Fourier mode of frequency $\omega$, and spatial momentum $q$ in the $x$-boundary direction. We now move on to grouping the different perturbations into gauge invariant ones that decouple from one another depending on the way they transform under the remaining rotational symmetry around the $x$-axis. The instabilities will arise in the longitudinal channel, where sound modes live, so we will only focus on such channel. The corresponding gauge invariant perturbations are
\begin{equation}
    \begin{aligned}
        Z_2 & = q^2h_{tt} + 2q\omega h_{tx}+\omega^2 h_{xx}-\frac{h_{yy}+h_{zz}}{2}\left[\omega^2+q^2\left(\frac{rg(r)f'(r)}{2g(r)-rg'(r)}-f(r)\right)\right],\\
        Z_{\psi} & = \delta\psi + \frac{h_{yy}+h_{zz}}{2}\frac{r^3\psi'(r)}{2g(r)-rg'(r)},
    \end{aligned}
\end{equation}
for which one can obtain a set of two coupled equations by combining the remaining equations until all gauge-dependent perturbations have been eliminated. We will omit these equations, as they are rather lengthy and do not offer much insight.

Discretizing the radial coordinate on a Chebyshev grid turns the resulting equations into a generalized eigenvalue problem that can be readily solved numerically~\cite{Jansen:2017oag}. To carry out these computations efficiently, we have developed \texttt{QUASTELL} (QUAsinormal Spectra of ThErmaL hoLography)~\cite{QUASTELL}, a \texttt{Julia} package that computes both the equilibrium background and its QNM spectrum using Chebyshev spectral methods, following the approach of~\cite{Jansen:2017oag}. The code is designed to be modular, so that the main ingredients of the model, such as the strength of the backreaction, the scalar potential, or even the equations of motion themselves, can be easily modified. The repository also includes \textit{Mathematica} notebooks that implement the same procedure for both the background and the QNM spectrum. We refer the reader to the repository~\cite{QUASTELL} for further details.

\begin{figure}[t!]
    \centering
    \includegraphics[width=0.49\linewidth]{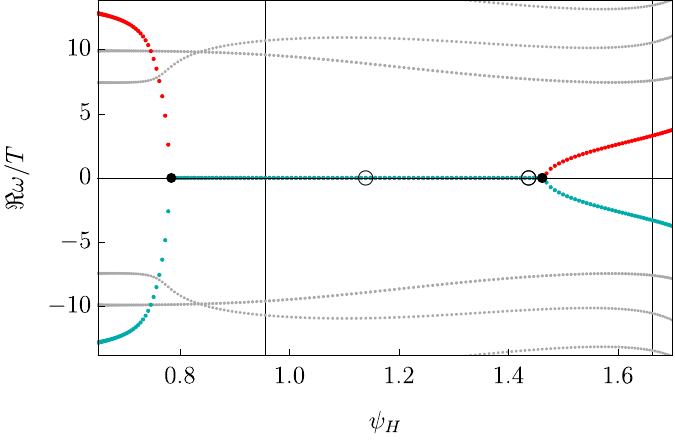}\hfill
    \includegraphics[width=0.49\linewidth]{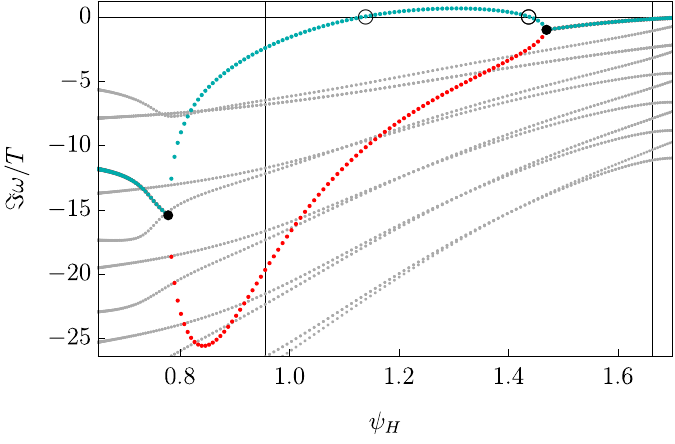}
    \caption{Real (left) and imaginary (right) part of the quasinormal mode frequencies at $q=0$ for theory A, as a function of the background scalar field value at the horizon. Highlighted with red and cyan are the non-hydrodynamic QNMs that we match to the effective theory. The small black dots correspond to additional gapped modes, not captured by the effective theory. The larger solid and empty dots show where the modes collide and where the lightest mode stability changes, respectively.}
    \label{fig:gapped-QNMs-q=0}
\end{figure}

\begin{figure}[t!]
    \centering
    \includegraphics[width=0.49\linewidth]{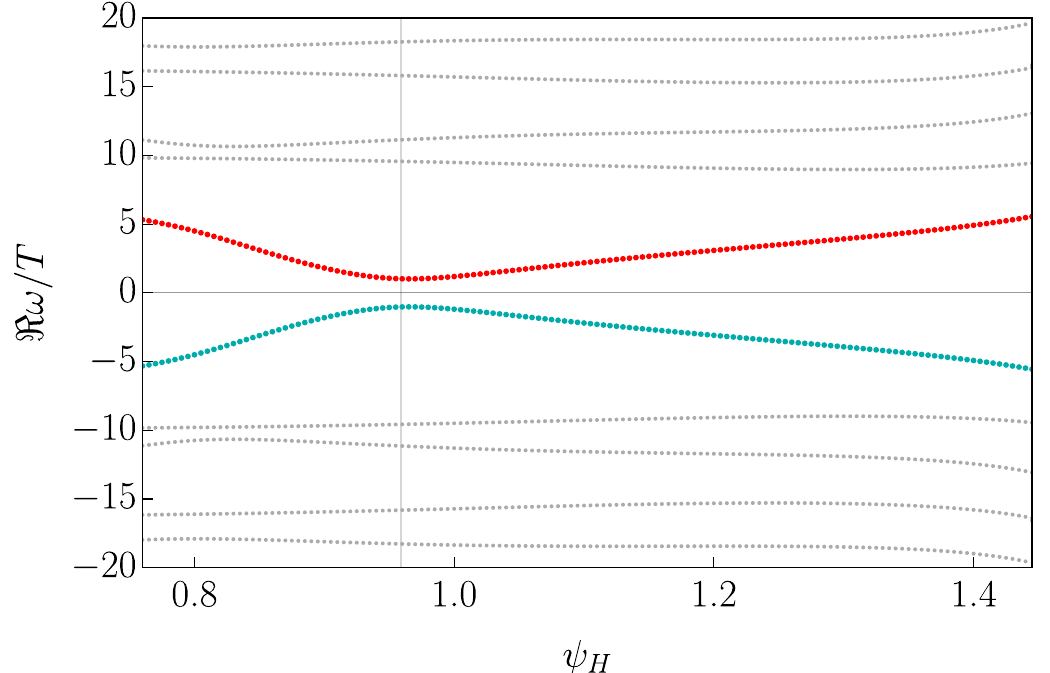}\hfill
    \includegraphics[width=0.49\linewidth]{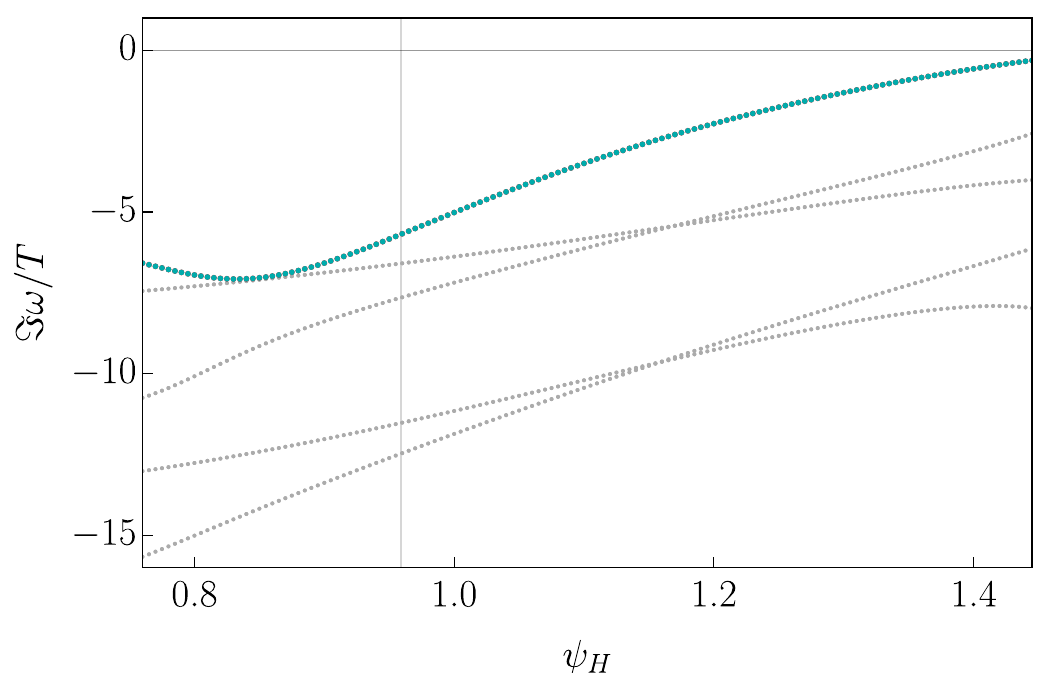}
    \caption{Real (left) and imaginary (right) part of the quasinormal mode frequencies at $q=0$ for theory B, as a function of the background scalar field value at the horizon. Highlighted with red and cyan are the non-hydrodynamic QNMs that we match to the effective theory. They exhibit the same imaginary part.
    The small black dots correspond to additional gapped modes, not captured by the effective theory.}
    \label{fig:gapped-QNMs-q=0-theoryB}
\end{figure}

We want to compare the holographic model with the effective theory presented in section~\ref{sec:scalar}.  The latter features two sound modes and two gapped modes, meaning modes with non-zero (complex) frequency at zero spatial momentum. On the other hand, the (longitudinal sector of the) holographic theory exhibits a richer spectrum with many gapped modes. The non-hydrodynamic modes at zero momentum are shown in red and cyan in Fig.~\ref{fig:gapped-QNMs-q=0} (theory A) and Fig.~\ref{fig:gapped-QNMs-q=0-theoryB} (theory B) as a function of the scalar field value at the horizon, $\psi_H$. For theory A, the two non-hydrodynamic modes matching the effective theory  can be selected and tracked as follows: In an intermediate interval of $\psi_H$ (whose endpoints are displayed with two big black dots), they are the only two purely imaginary modes. At the endpoints, these two gapped modes merge. Outside the interval, they form a pair of modes with the same imaginary part and opposite real part. For theory B, we extract them at some $\psi_H$ inside the spinodal branch (between the vertical gray lines in Fig.~\ref{fig:gapped-QNMs-q=0-theoryB}), where they are the lightest, and then we track them backwards and forward in $\psi_H$.

\subsection{Extracting the effective theory coefficients}
\label{sec:extractingEFTcoeffs}

Solving for the modes in the dispersion relation \eqref{eq:dispersion} requires knowledge of the effective field theory coefficients. More specifically, we only need to determine the combinations in \eqref{eq: EFT parameters}. For this purpose, we use the quasinormal mode spectrum in specific regimes. 

\begin{figure}[t]
    \centering
   \includegraphics[width=0.49\linewidth]{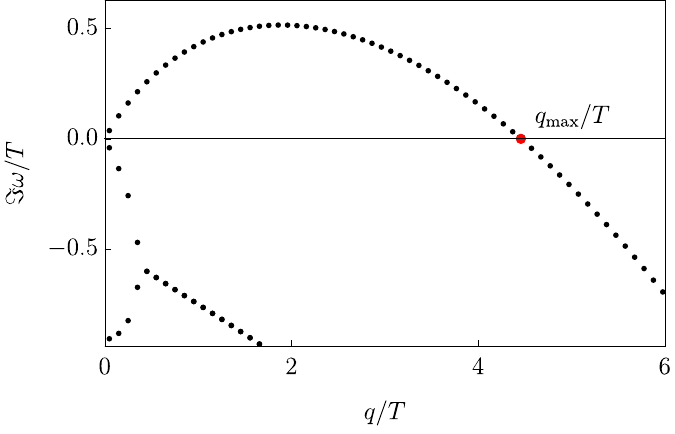}
    \includegraphics[width=0.49\linewidth]{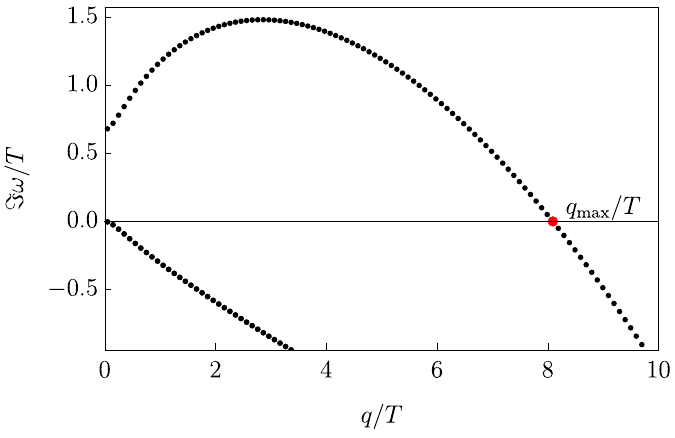}
    \caption{Dispersion relation of the unstable modes in the spinodal branch for background 3 (left) where sound is unstable, and background 4 (right) where the gapped mode is unstable. The crossing point $q_{\text{max}}$ can be used to obtain effective theory parameters.}
    \label{fig:q02-in-QNMs}
\end{figure}

\begin{figure}[t]
    \centering
    \includegraphics[width=0.55\linewidth]{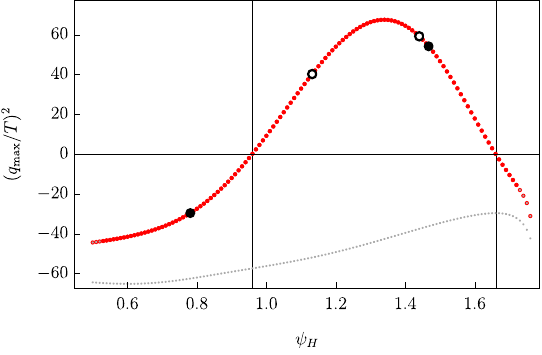}
    \caption{Value of $(q_{\text{max}}/T)^2$ as a function of the scalar field at the horizon $\psi_H$. Vertical black lines represent the boundary of the spinodal branch. Empty dots indicate the instability-switching points, whereas big black dots represent the points at which the gapped modes merge.  The small black dots correspond to an additional zero-frequency mode, not captured by the effective theory.}
    \label{fig:qmaxtheoryA}
\end{figure}

First, the effective theory predicts that the frequency of a mode vanishes at a finite momentum \eqref{eq:momentumdisp}. This is in agreement with the microscopic theory calculation of quasinormal modes, as can be observed in Fig.~\ref{fig:q02-in-QNMs}, where we show the $q$-dependence  of the lightest quasinormal modes for two states in the spinodal branch. In the left panel, the sound mode is unstable, whereas in the right panel the instability arises from the gapped mode. Hence, depending on the type of instability, a different mode has vanishing frequency at nonzero momentum. By tracking the vanishing frequency point in Fig.~\ref{fig:q02-in-QNMs} for different background states, we can extract the coefficient \eqref{eq:momentumdisp}. A plot of $q_{\text{max}}^2$ as a function of the scalar field at the horizon is provided in Fig.~\ref{fig:qmaxtheoryA} for theory A. There, one can observe how this quantity remains positive as long as the state is in the spinodal branch. In this section, all the quantities are plotted for theory A, which exhibits the interplay of instabilities. Plots for theory B, where the instability is always triggered by the sound mode, are collected in appendix~\ref{app:TheoryBplots}.

\begin{figure}[t]
    \centering
    \includegraphics[width=0.48\linewidth]{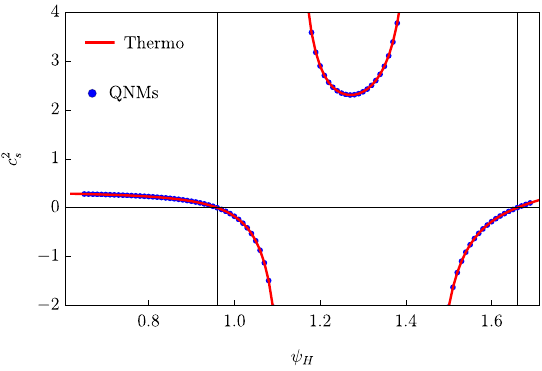}
    \hfill
    \includegraphics[width=0.48\linewidth]{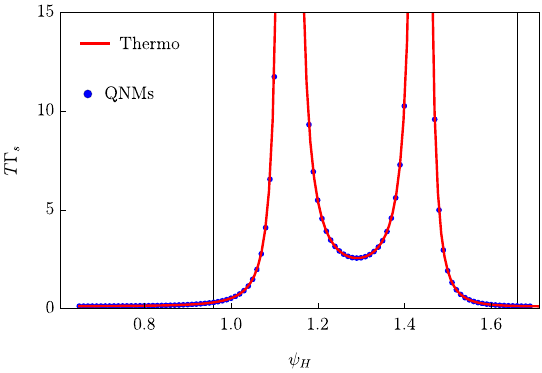}
    
    \caption{Speed of sound squared (left) and sound attenuation constant $\Gamma_s$ (right) as a function of the scalar field at the horizon for theory A. The red lines are obtained from the background solutions using the relations (\ref{eq:cs2fromthermo}) (left) and (\ref{eq:GammaS}) (right), whereas the blue dots are computed from the sound QNMs, namely (\ref{eq:alphasound1}) (left) and (\ref{eq:alphasound2}) (right). Vertical black lines represent the boundary of the spinodal branch.}
    \label{fig:cs2andGammascomparison}
\end{figure}

Next, we turn to the low-momentum regime. Introducing the notation 
\begin{eqnarray}
\label{eq:M1andM2}
    M_1 = \frac{\partial_\phi^2V}{Z_0} \frac{c_v}{C} \ ,\quad \quad \quad M_2 = \frac{\gamma}{2Z_0} \ , \quad \quad \quad \Delta = \sqrt{M_1 - M_2^2} \ ,  
\end{eqnarray}
the dispersion relations (\ref{eq:dispersionGapMode}) and (\ref{eq:dispersionSound2}) for the gapped  and sound modes read
\begin{subequations}
    \begin{eqnarray}
        \omega_\pm(q)&=&\alpha_{\pm,0}+\alpha_{\pm,2}q^2+ O(q^4) \ , \\
        \omega_s(q) &=& \pm\alpha_{s,1}q+\alpha_{s,2}q^2+ O(q^3) \ , 
    \end{eqnarray}
\end{subequations}
where
\begin{subequations}
    \begin{eqnarray}
    \alpha_{\pm,0} &=& -i M_2\pm \Delta \ ,  \\
\label{eq:alphapm2b}
    \alpha_{\pm,2} &=& \frac{
\left(\frac{c_v}{C}-1\right)c_s^2 M_1
+ \frac{Z_s}{Z_0} \left(
\mp 2iM_2\sqrt{M_1-M_2^2}
+M_1-2M_2^2
\right)
}{
2M_1\left(\pm \sqrt{M_1-M_2^2}-2iM_2\right)
-4M_2^2\left( \pm \sqrt{M_1-M_2^2}-iM_2\right)
} \ , \\
\label{eq:alphasound1}
    \alpha_{s,1} &=& c_s=\sqrt{\frac{s}{c_v}} \ ,\\
    \label{eq:alphasound2}
    \alpha_{s,2} &=& -\frac{i}{2} \left[\Gamma -2c_s^2\frac{M_2}{M_1}\left(1-\frac{c_v}{C}\right)\right]=-\frac{i}{2}\Gamma_s \ .
 \end{eqnarray}
\end{subequations}
From the linear and quadratic terms in the sound mode dispersion relation, we directly extract $c_s^2=s/c_v$ and $\Gamma_s$, respectively. These quantities can also be computed from the background solutions using the thermodynamic relations (\ref{eq:cs2fromthermo}) and (\ref{eq:GammaS}). Fig.~\ref{fig:cs2andGammascomparison} compares the results obtained from the background solutions (red lines) with those extracted from the QNMs (blue dots), showing agreement within numerical precision. Both quantities diverge at $\psi_H \approx 1.13$ and $\psi_H \approx 1.44$, where the instability shifts from one kind of mode to the other.

\begin{figure}[t!]
    \centering
  
  \includegraphics[width=0.46\linewidth]{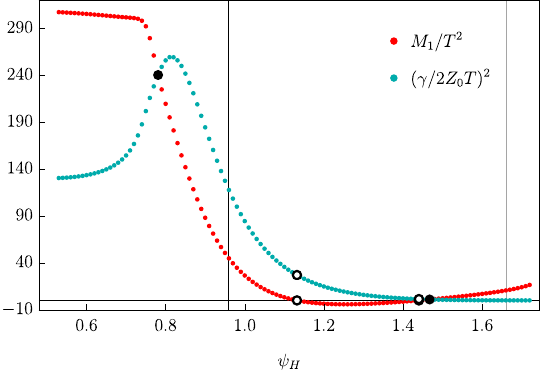}
  \hspace{2mm}  \includegraphics[width=0.49\linewidth]{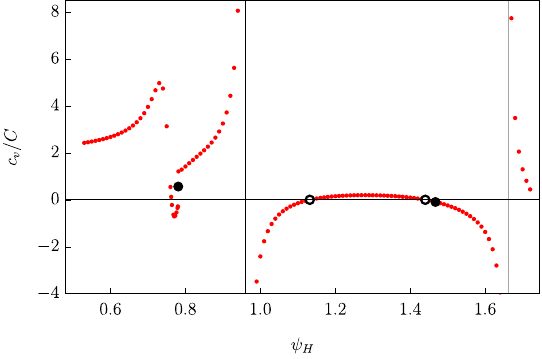}\hfill
    \includegraphics[width=0.465\linewidth]{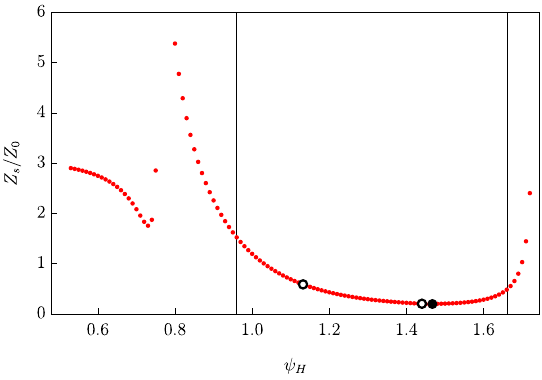}\hfill
    \includegraphics[width=0.49\linewidth]{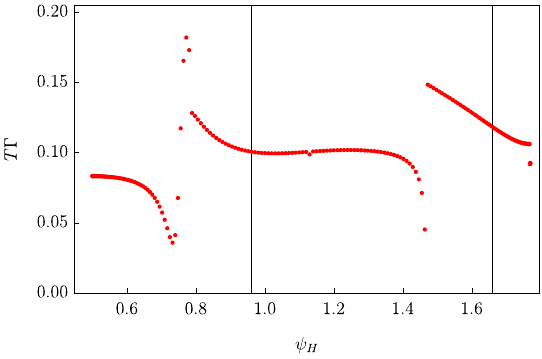}\hfill

    \caption{Effective theory parameters as functions of the scalar field at the horizon for theory A: $M_1$ and $M_2^2 = \left( \gamma/2 Z_0\right) ^2$ (top-left), $c_v/C$ (top-right), $Z_s/Z_0$ (bottom-left) and $\Gamma$ (bottom-right). All dimensionful quantities are expressed in units of temperature.   Vertical black lines represent the boundary of the spinodal branch. Empty dots indicate the instability-switching points, whereas black dots represent the points at which the gapped modes merge.}
    \label{fig:EFT-parameters}
\end{figure}

From the zero-momentum term of the gapped modes, since $M_2$ is real, we obtain
\begin{equation}
    M_2= \frac{\gamma}{2 Z_0} =  -{\rm Im}\left(\frac{\alpha_{+,0}+\alpha_{-,0}}{2}\right) \ .
\end{equation}
On the other hand, $\Delta$ can be either real or purely imaginary. In the example shown in Fig.~\ref{fig:gapped-QNMs-q=0}, where $\alpha_{\pm,0}$ are plotted, the frequencies are purely imaginary in the region between the two black dots, implying that $\Delta$ is purely imaginary there. Outside this region, $\alpha_{\pm,0}$ acquire a nonzero real part, indicating that $\Delta$ is real. Hence, 
\begin{itemize}
\item If $\Delta$ is real
\begin{equation}
M_1= M_2^2+   \left[{\rm Re}\left(\frac{\alpha_{+,0} -\alpha_{-,0}}{2}\right)\right]^2,
\end{equation}
\item If $\Delta$ is purely imaginary
\begin{equation}
M_1=M_2^2- \left[{\rm Im}\left(\frac{\alpha_{+,0} -\alpha_{-,0}}{2}\right)\right]^2.
\end{equation}
\end{itemize}
A plot of $M_1/T^2$ and $(M_2/T)^2=(\gamma/2Z_0T)^2$ is presented in the top-left panel of Fig.~\ref{fig:EFT-parameters}. Recalling (\ref{eq:M1andM2}), the two curves cross when $\Delta = 0$. This indeed occurs at the $\psi_H$ values for which the modes merge, as  we can see in Fig.~\ref{fig:gapped-QNMs-q=0}. We also see that $M_1$ is negative when the lightest gapped mode is unstable. 

Once $M_1$ and $M_2$ are known, we can use the expression (\ref{eq:alphapm2b}) for $\alpha_{\pm,2}$ to determine $c_v/C$ as follows:
\begin{itemize}
    \item If $\Delta$ is real
\begin{equation}\label{eq:kvsol}
    \frac{c_v}{C} = 1+\frac{M_1}{c_s^2 M_2}{\rm Im}\,\alpha_{+,2} \ .
\end{equation}
\item If $\Delta=i\tilde \Delta=i\sqrt{M_2^2-M_1}$ is purely imaginary, we use (\ref{eq:momentumdisp}) together with $M_1$ to eliminate $Z_s/Z_0$ from (\ref{eq:alphapm2b}), obtaining a quadratic equation for $c_v/C$:
\begin{equation}
  - 2 {\rm Im}\alpha_{+,2}=\frac{M_1}{q_{\text{max}}^2} \frac{C}{c_v}\frac{M_1-2M_2^2+2 M_2\tilde \Delta +c_s^2 q_{\text{max}}^2 \frac{c_v}{C}(1-\frac{c_v}{C})}{M_1(2 M_2-\tilde \Delta)+2 M_2^2(\tilde \Delta-M_2)}\ .
\end{equation}
The solutions are of the form
\begin{equation}\label{eq:kvsol2}
    \frac{c_v}{C}= \frac{1}{2}+\frac{{\rm Im}\,\alpha_{+,2}}{\mathcal{B} \mathcal{P}}\pm \sqrt{\frac{\mathcal{A}}{\mathcal{B}}+\left( \frac{1}{2}+\frac{{\rm Im}\,\alpha_{+,2}}{\mathcal{B} \mathcal{P}}\right)^2}\ ,
\end{equation}
where
\begin{equation}
    \mathcal{A} = - \left( \tilde \Delta - M_2 \right)^2 \ ,\quad \quad \quad \mathcal{B} =c_s^2 q_{\text{max}}^2 \ 
\end{equation}    
and
\begin{equation}
\mathcal{P} =\frac{M_1}{q_{\text{max}}^2\left(M_1(2 M_2-\tilde \Delta)+2 M_2^2(\tilde \Delta-M_2)\right)}\,.
\end{equation}
We choose the root that yields a positive $c_v/C$ in the region where the sound modes are stable and a negative $c_v/C$ when they are not.

\end{itemize}
Given $c_v/C$, $Z_s/Z_0$ is readily found from
\be
\label{eq:alternativez}
\frac{Z_s}{Z_0} = - \frac{M_1}{q_{\text{max}}^2} \frac{C}{c_v} \ .
\ee
In Fig.~\ref{fig:EFT-parameters}, we plot $c_v/C$ in the top-right panel and $Z_s/Z_0$ in the bottom-left panel. We see that $c_v/C$ is positive when the sound modes are stable, except for some points around $\psi_H \approx 0.77$. There, we observe a small region of negative values, followed by a discontinuity, even though the corresponding values of $\psi_H$ lie on the metastable branch. Around the same values we observe that $Z_s/Z_0$ has a pronounced peak. These unexpected features most likely arise because this range of $\psi_H$ lies outside the regime of validity of the effective theory. Indeed, Fig.~\ref{fig:gapped-QNMs-q=0} shows that, at these values of $\psi_H$, the spectrum contains gapped modes (shown as black dots) that are lighter than the gapped mode described by the effective theory (shown in cyan). We will discuss in more detail the validity of the effective theory in section~\ref{sec:validityofEFT}.

With the computations described so far, we are able to find all the quantities in (\ref{eq: EFT parameters}) except $\Gamma$. This can be obtained by numerically extracting the quadratic term of the sound modes. Indeed, from (\ref{eq:alphasound2}), we find 
\be
\label{eq:Gammafrommodes}
     \Gamma = \Gamma_s+2c_s^2\frac{M_2}{M_1}\left(1-\frac{c_v}{C}\right) \ .
\ee
A plot of $\Gamma$ as a function of the scalar field at the horizon is presented in the bottom-right panel of Fig.~\ref{fig:EFT-parameters}.

\subsection{Validity of the effective theory}
\label{sec:validityofEFT}

Now that we have extracted all coefficient combinations that are required to obtain the effective theory mode behavior, we will test how accurate it is in capturing the dispersion relation of the modes of the full holographic description. Using the notation introduced in the previous section, \eqref{eq:dispersion} takes the form
\begin{equation}
\begin{aligned}
        \omega^4 & +i\omega^3\left( 2M_2+\Gamma q^2\right)-\omega^2\left[M_1+\left(\frac{c_v}{C}c_s^2+2M_2 \Gamma+\frac{Z_s}{Z_0}\right) q^2\right]\\
          & -i\omega q^2\left[2c_s^2 M_2\frac{c_v}{C}+\Gamma\left(M_1+ q^2\frac{Z_s}{Z_0}\right)\right]+c_s^2 q^2\left(M_1+\frac{c_v}{C}\frac{Z_s}{Z_0} q^2\right) = 0 \ .
\end{aligned}
\label{eq:dispersion-2}
\end{equation}
As the effective theory has been built as a series expansion in gradients, strictly speaking we would need to solve \eqref{eq:dispersion-2} perturbatively in momenta, which we have already done to fit several of the effective theory parameter combinations. Nonetheless, one can always solve \eqref{eq:dispersion-2} for any value of $q/T$. In what follows, we show that the quasinormal mode behavior is typically well captured by the effective theory even outside the small momentum regime. Moreover, we will see that the effective theory improves, in many cases significantly,  on the results of the purely hydrodynamic description.

\begin{figure}
    \centering
    \includegraphics[width=0.49\linewidth]{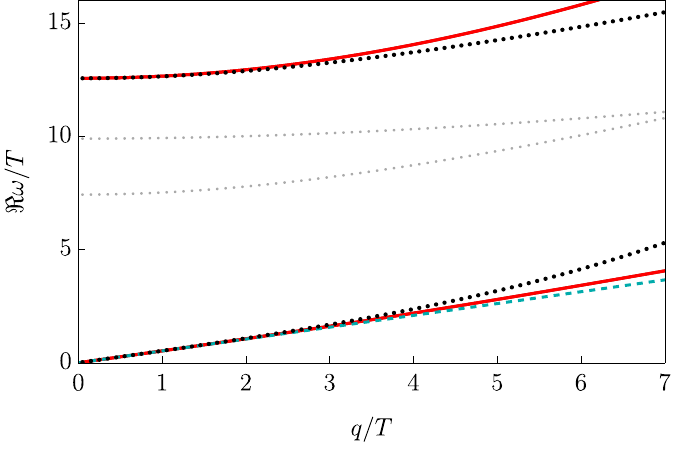}
    \hfill
    \includegraphics[width=0.49\linewidth]{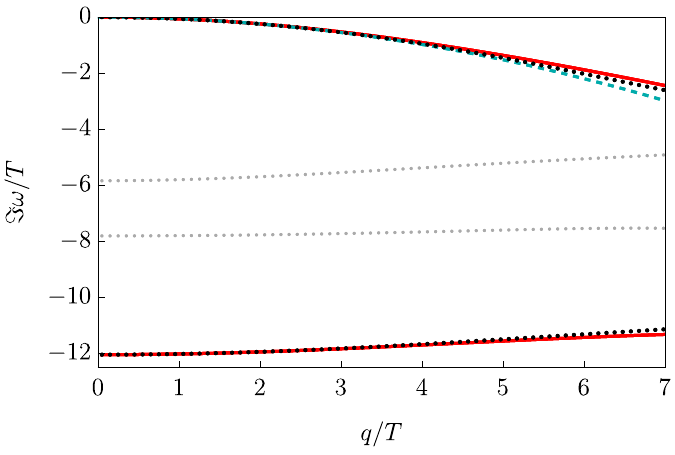}
    
    \includegraphics[width=0.49\linewidth]{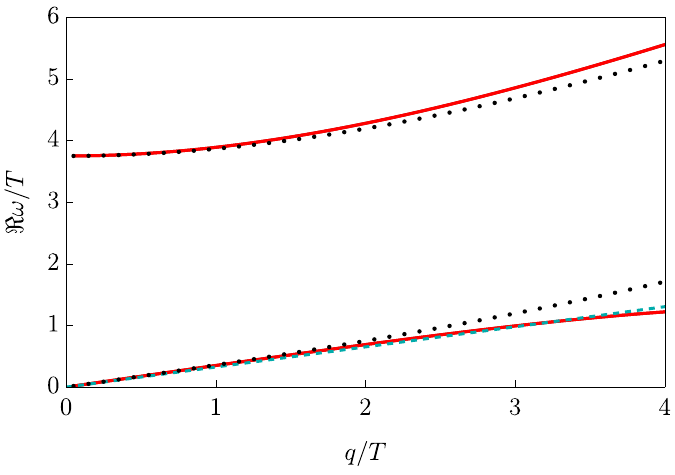}
    \hfill
    \includegraphics[width=0.49\linewidth]{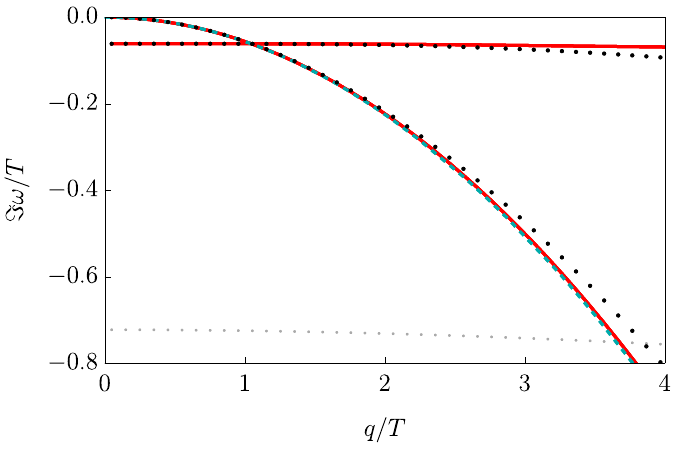}
    \caption{Effective theory (in solid red) and quasinormal mode (black dots) comparison for theory A for backgrounds (top) 1 and (bottom) 6. For reference, we also include the hydrodynamic dispersion relation in dashed cyan and the rest of the gapped modes as gray dots.}
    \label{fig:QNM-vs-EFT-A-1}
\end{figure}

\begin{figure}
    \centering
    \includegraphics[width=0.49\linewidth]{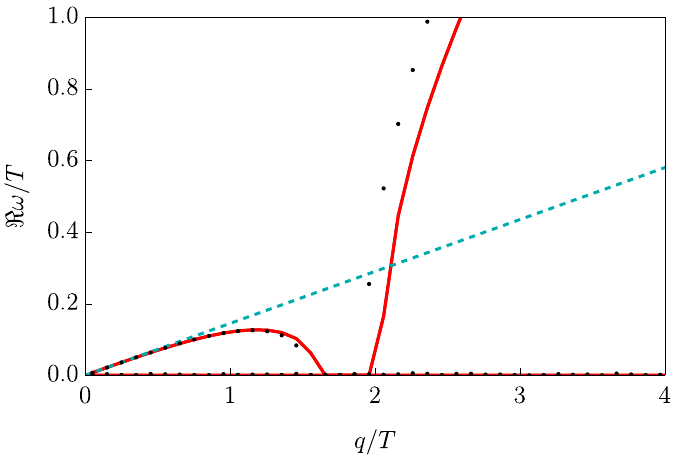}
    \hfill
    \includegraphics[width=0.49\linewidth]{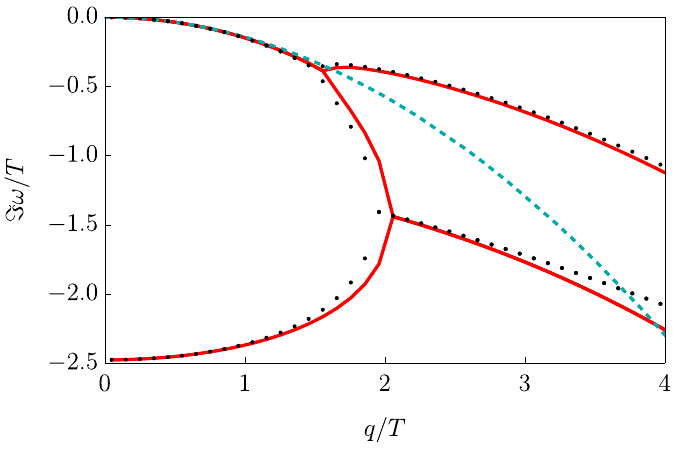}
    
    \caption{Effective theory (in solid red) and quasinormal mode (black dots) comparison for theory A for background 2. For reference, we also include the hydrodynamic dispersion relation in dashed cyan and the rest of the gapped modes as gray dots.}
    \label{fig:QNM-vs-EFT-A-2}
\end{figure}

\begin{figure}
    \centering
    \includegraphics[width=0.49\linewidth]{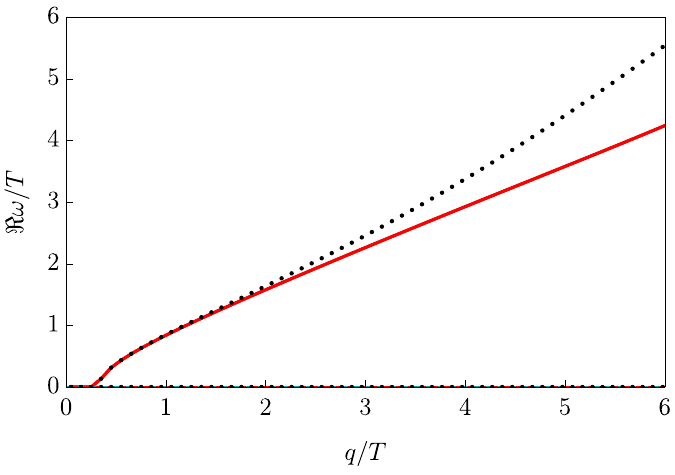}
    \hfill
    \includegraphics[width=0.49\linewidth]{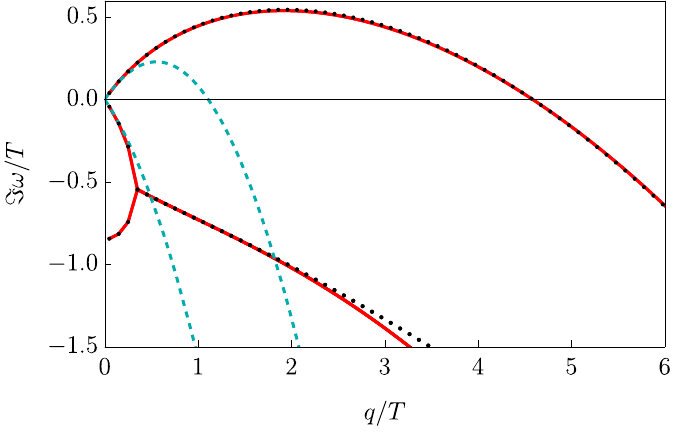}
    
    \includegraphics[width=0.49\linewidth]{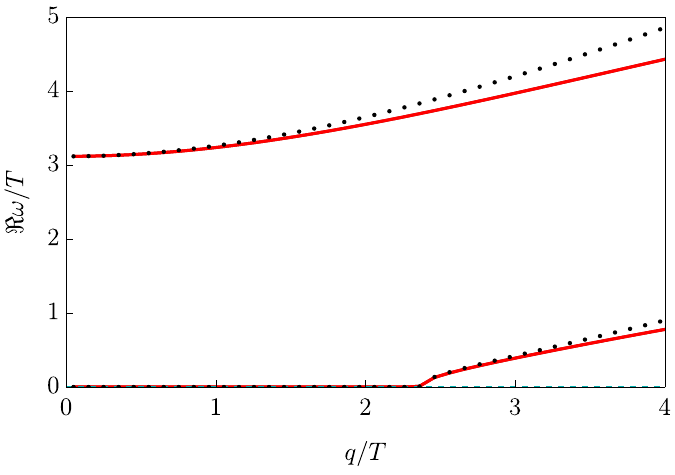}
    \hfill
    \includegraphics[width=0.49\linewidth]{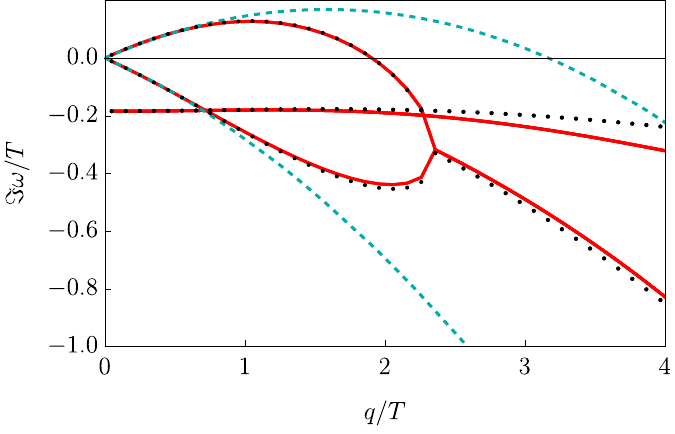}
    \caption{Effective theory (in solid red) and quasinormal mode (black dots) comparison for theory A for backgrounds (top) 3 and (bottom) 5. For reference, we also include the hydrodynamic dispersion relation in dashed cyan and the rest of the gapped modes as gray dots.}
    \label{fig:QNM-vs-EFT-A-3}
\end{figure}

\begin{figure}
    \centering
    \includegraphics[width=0.49\linewidth]{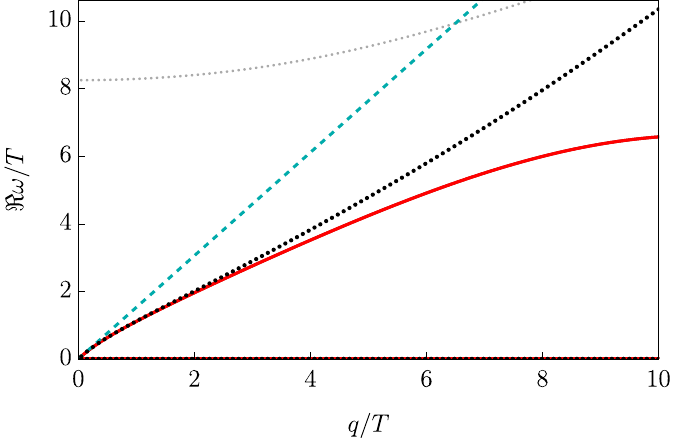}
    \hfill
    \includegraphics[width=0.49\linewidth]{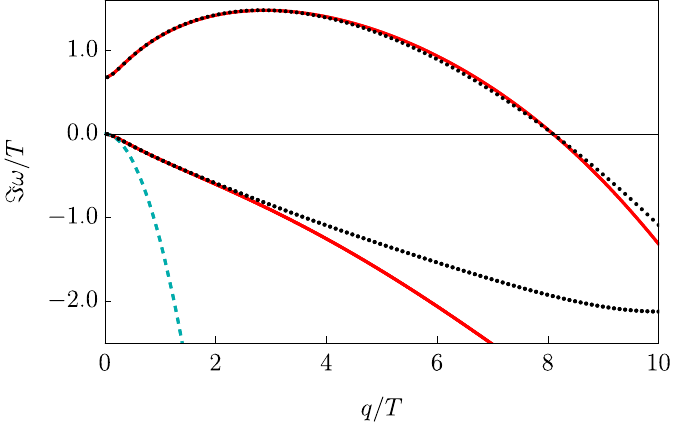}
    
    \caption{Effective theory (in solid red) and quasinormal mode (black dots) comparison for theory A for background 4. For reference, we also include the hydrodynamic dispersion relation in dashed cyan and the rest of the gapped modes as gray dots.}
    \label{fig:QNM-vs-EFT-A-4}
\end{figure}

\begin{figure}
    \centering
    \includegraphics[width=0.49\linewidth]{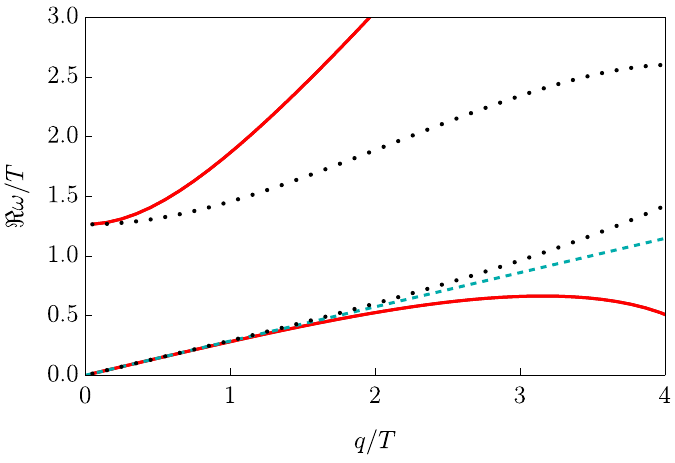}
    \hfill
    \includegraphics[width=0.49\linewidth]{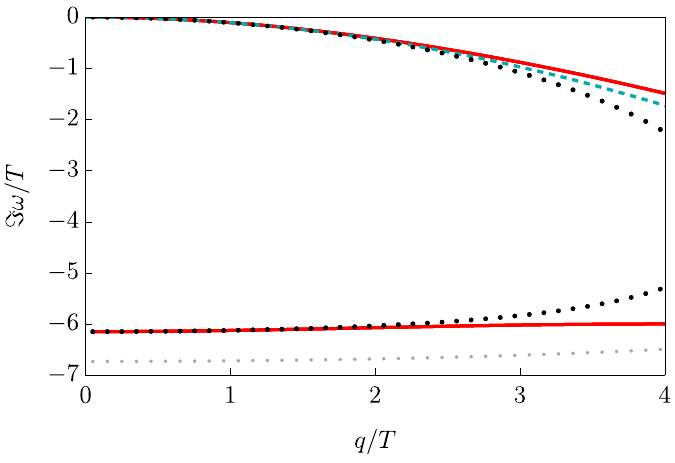}
    
    \includegraphics[width=0.49\linewidth]{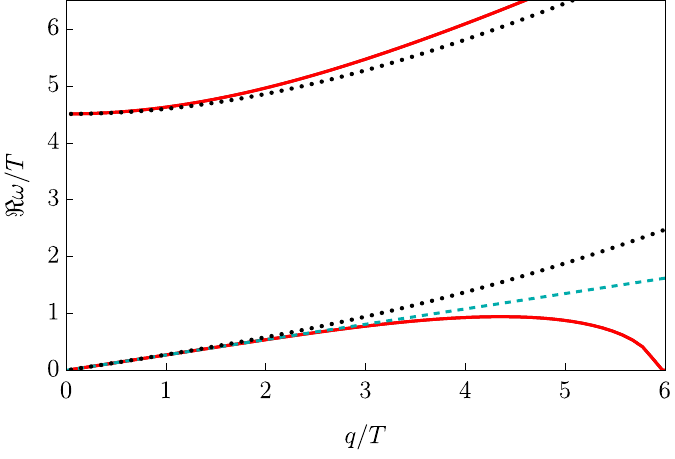}
    \hfill
    \includegraphics[width=0.49\linewidth]{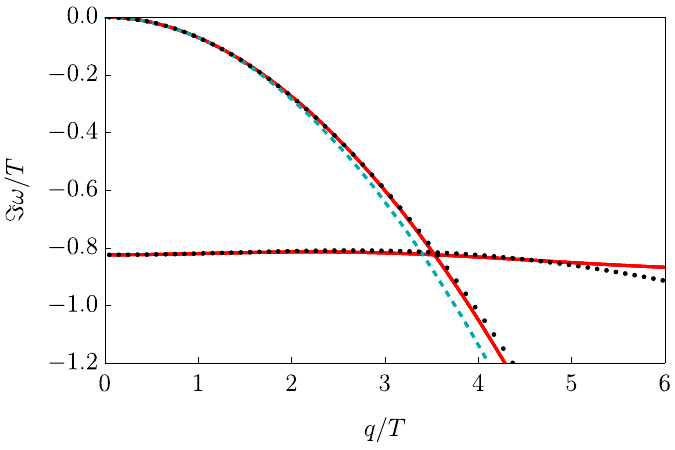}
    \caption{Effective theory (in solid red) and quasinormal mode (black dots) comparison for theory B for backgrounds (top) 1 and (bottom) 5. For reference, we also include the hydrodynamic dispersion relation in dashed cyan and the rest of the gapped modes as gray dots.}
    \label{fig:QNM-vs-EFT-B-1}
\end{figure}

\begin{figure}
    \centering
    \includegraphics[width=0.49\linewidth]{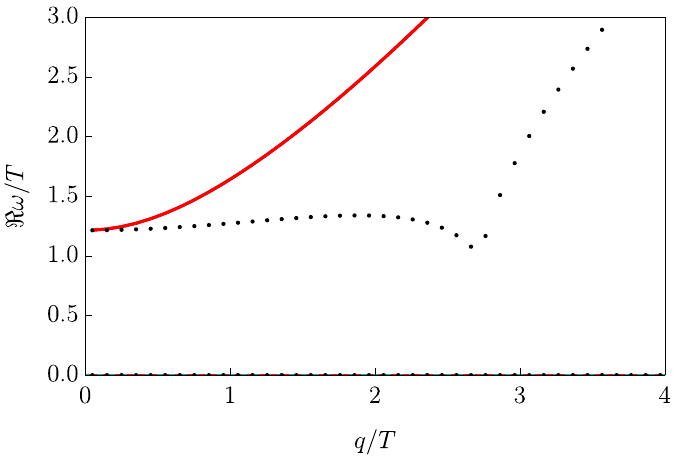}
    \hfill
    \includegraphics[width=0.49\linewidth]{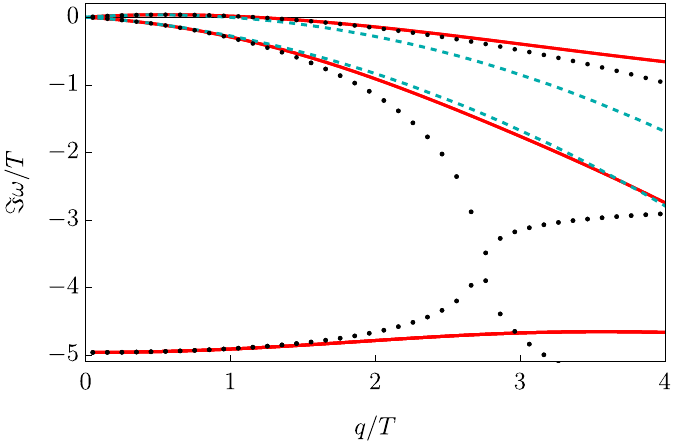}
    
    \includegraphics[width=0.49\linewidth]{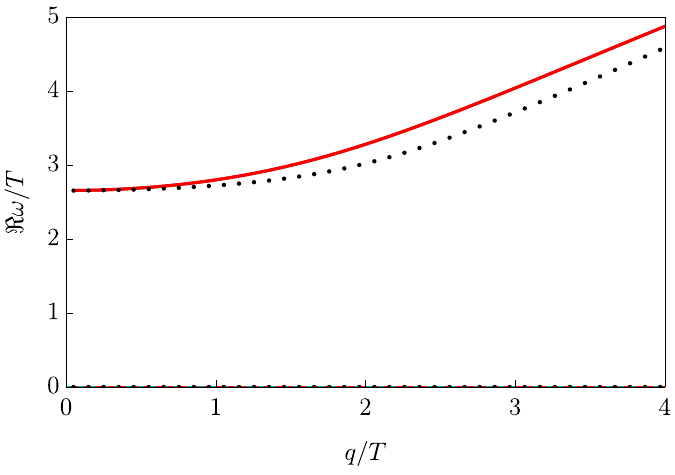}
    \hfill
    \includegraphics[width=0.49\linewidth]{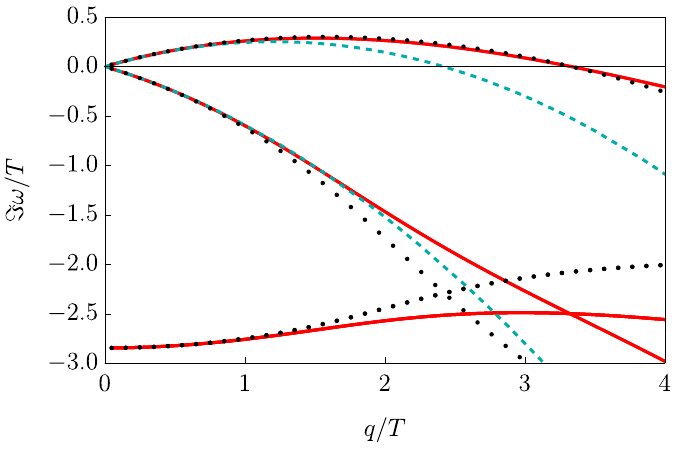}

    \includegraphics[width=0.49\linewidth]{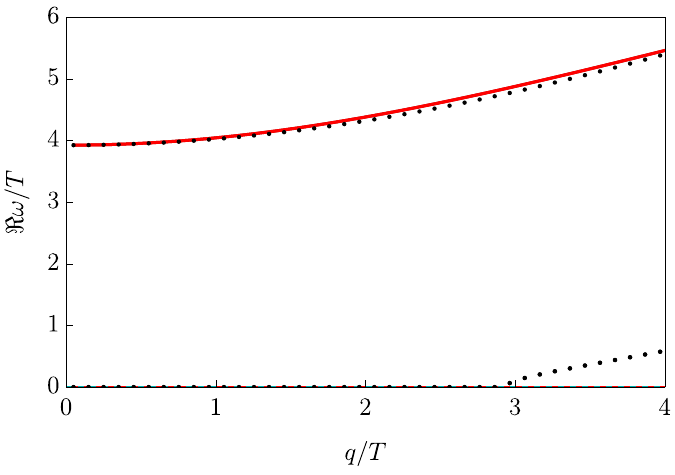}
    \hfill
    \includegraphics[width=0.49\linewidth]{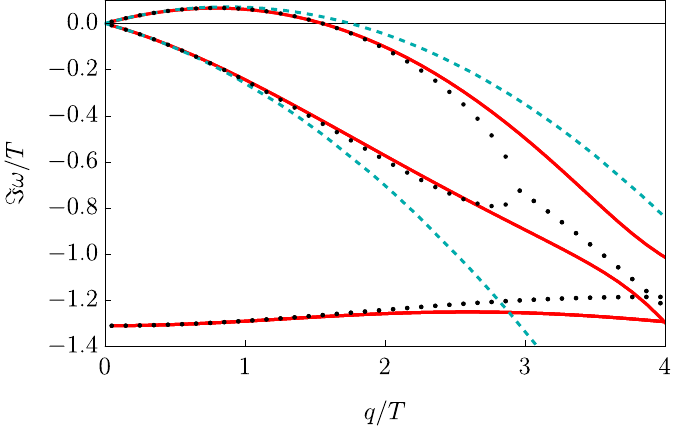}
    \caption{Effective theory (in solid red) and quasinormal mode (black dots) comparison for the theory B for backgrounds (top) 2 (middle) 3 and (bottom) 4. For reference, we also include the hydrodynamic dispersion relation in dashed cyan and the rest of the gapped modes as gray dots.}
    \label{fig:QNM-vs-EFT-B-2}
\end{figure}

We display results for the backgrounds shown in Fig.~\ref{fig:s-of-T}, which together cover all the qualitatively distinct behaviors of the relevant quasinormal modes that we observe. We show the result of the comparison between the effective theory and quasinormal modes in Fig.~\ref{fig:QNM-vs-EFT-A-1}--\ref{fig:QNM-vs-EFT-A-4} for theory A, and in Fig.~\ref{fig:QNM-vs-EFT-B-1}--\ref{fig:QNM-vs-EFT-B-2} for theory B. We represent as solid red lines the solutions to \eqref{eq:dispersion-2} and as black dots the QNM frequencies of the relevant modes. Additionally, we have included the sound dispersion relation \eqref{eq:dispersionSound2} in dashed cyan and the rest of the gapped modes as gray dots.

Let us focus first on theory A. Fig.~\ref{fig:QNM-vs-EFT-A-1} shows the comparison for backgrounds 1 and 6, which are locally stable and whose gapped modes have the same imaginary part of the frequency and opposite real parts. We observe that, despite the gapped modes being quite heavy, the effective theory dispersion captures the right behavior well up to $q\sim 4T$, where deviations become significant, presumably due to a combination of high momenta and the modes approaching other gapped ones. The purely hydrodynamic sound dispersion relation follows closely the gapless mode of the effective theory in the range of momenta where they accurately describe the QNM data. 

In Fig.~\ref{fig:QNM-vs-EFT-A-2}, we show the comparison for background 2, which is also in a locally stable branch, but whose gapped modes are purely imaginary. From both gapped modes, we only discuss the lightest; the heavier one is not well captured by the effective theory as it is heavier than other gapped modes, see Fig.~\ref{fig:gapped-QNMs-q=0}. The speed of sound is small due to its proximity to the spinodal point. Once again, the effective theory modes describe well the QNM data until $q\sim 4T$, but this time the mode behavior is richer, with different collisions taking place. At around $q \sim T$, both sound modes collide, at which point the purely hydrodynamic sound dispersion relation fails to follow correctly the data. The modes predicted by the improved effective theory, however, capture well this collision. Later, at $q\sim 2T$, the heaviest sound mode collides with the lightest gapped mode, which results in two modes with the same imaginary part and opposite real parts that get further apart with increasing momentum. As can be seen, this second collision is also captured by our effective theory.

We emphasize that, apart from the value of $q_{\text{max}}$, the parameters were fixed using data in a range of very small momentum, up to $q\sim 10^{-3}T$. Nonetheless, the effective theory captures qualitative and often quantitative features of the spectrum at $q\sim T$, including mode collisions.

Turning now to Fig.~\ref{fig:QNM-vs-EFT-A-3}, we show the comparison for backgrounds 3 and 5. These states lie on the branch of the spinodal where sound is unstable, but in the first one the relevant gapped modes are purely imaginary. Both plots exhibit collisions that are well captured by the effective theory. In the first case, the collision is between the stable sound mode and the lightest gapped one; while in the second case, the collision takes place between both sound modes. The stability crossing point $q_{\text{max}}$ is used to fit parameters but the unstable mode is very well captured by the EFT in the entire $0<q<q_{\max}$ range, and even beyond for the first background.

Finally, we have the results for background 4 in Fig.~\ref{fig:QNM-vs-EFT-A-4}. This state is locally unstable, but the instability is that of a gapped mode. The effective theory captures the behavior in a range of momenta up to $q\sim 2 T$, and again fares considerably better than the purely hydrodynamic sound dispersion relation. Contrary to the previous cases, even if the location $q_{\text{max}}$ is accurate (by construction), the unstable mode behavior deviates  more noticeably from the full holographic result for $q\gtrsim 2T$. An observation is that, in this case, $q_{\max}$ is around the largest it can be for this theory, and there is another mode at zero frequency with $|q^2|<q_{\max}^2$ (see Fig.~\ref{fig:qmaxtheoryA}).  If the background is closer to the spinodal point, both the frequency of the gapped mode and $q_{\max}$ decrease, leading to a better agreement with the QNM data. 

We now move on to theory B. Here, we can only distinguish between two main types of backgrounds, characterized by whether sound is stable or unstable.

In Fig.~\ref{fig:QNM-vs-EFT-B-1}, we show the comparison between effective and quasinormal modes for the backgrounds 1 and 5, which are locally stable. We observe that the effective theory captures the correct behavior of sound modes up to  $q\sim 2T$, similarly to the purely hydrodynamic sound dispersion relation, although the effective theory fares slightly better reproducing the imaginary part of the sound dispersion relation. For the gapped mode, there can be appreciable deviations in the dispersion relation at relatively low momentum when it is heavy, but the effective theory result improves when the gapped mode becomes lighter.

In Fig.~\ref{fig:QNM-vs-EFT-B-2}, we show the results for backgrounds 2 to 4, which are unstable. In this case the effective theory dispersion relation also follows the sound modes better than the purely hydrodynamic dispersion, although they deviate earlier than in the A theory. As in backgrounds 1 and 5, the effective theory reproduces the gapped mode more accurately as it becomes lighter.

\subsection{QNMs with varying backreaction}

In the probe scalar model of \cite{Caddeo:2024lfk}, the lowest QNM near the spinodal points is purely imaginary, with a decay rate proportional to the curvature of the effective potential and a damping that grows quadratically with momentum. The decay rate therefore vanishes at the spinodal points. Formally continued into the spinodal branch, where $\partial_\phi^2 V < 0$, this corresponds to a mode that is unstable at zero momentum. This matches the effective theory in the probe approximation. There, away from the spinodal points, the specific heat is determined by the thermal bath $c_v \approx C$, which is positive for an ordinary fluid. The temperature and velocity fluctuations are decoupled from the scalar and the sound modes remain stable, while a non-hydrodynamic mode of the scalar is unstable already at $q = 0$. 

\begin{figure}[t!]
    \centering
    \includegraphics[width=0.9\linewidth]{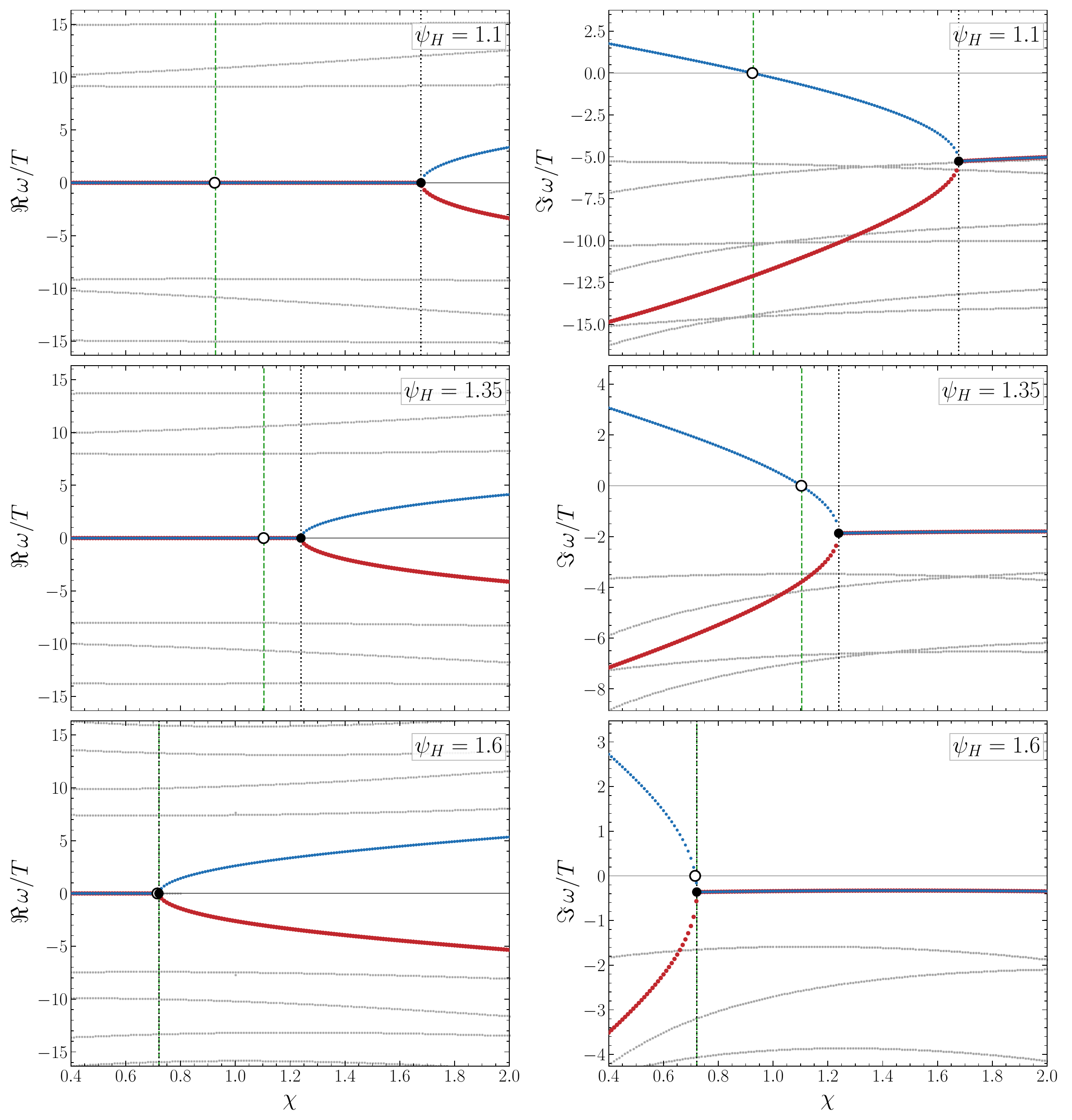}
    \caption{Real (left) and imaginary (right) parts of the QNM frequencies at $q = 0$ as a function of $\chi$, for the backgrounds marked with dashed gray lines in Fig.~\ref{fig:SofT_backreaction}. The unstable mode (blue) and its collision partner (red) are highlighted, and the rest of the spectrum is shown in gray. The white dot marks where the unstable mode becomes stable. It coincides with the crossing of the $ds/dT = 0$ curve (blue in Fig.~\ref{fig:SofT_backreaction}), indicated here by the green dashed vertical line. The black dot marks the collision of the two modes (black dotted line).}
    \label{fig:QNMs_backreaction}
\end{figure}

To see how this picture changes as the backreaction is switched on, we compute the QNMs at zero momentum along the three slices of constant $\psi_H$\footnote{Note that the temperature in units of the source $T/\Lambda$ will generically change along these slices, however this does not fundamentally affect the analysis.} marked in Fig.~\ref{fig:SofT_backreaction}. The results are shown in Fig.~\ref{fig:QNMs_backreaction}. In the three cases we start at small $\chi$ (small backreaction) inside the spinodal branch at a point where the sound mode is stable, $c_v>0$ and there is an unstable non-hydrodynamic mode. The two highlighted modes in Fig.~\ref{fig:QNMs_backreaction} show the expected behavior corresponding to the non-hydrodynamic modes of the effective theory. They are both purely imaginary, and one of them has a positive imaginary part, so it is unstable. As $\chi$ increases, this mode becomes stable (white dot). Note that the white dots lie on the $ds/dT = 0$ curve (vanishing specific heat) of Fig.~\ref{fig:SofT_backreaction}, which is computed from the equilibrium solutions alone. Immediately beyond the white dot the state is still on the spinodal branch, but now with $c_v < 0$, and the effective theory predicts that the instability is carried by a sound mode at finite momentum, which is consistent with the microscopic result. This confirms the expected crossover from non-hydrodynamic to hydrodynamic instability, as increasing the number of degrees of freedom involved in the transition changes the thermodynamic stability.

 If we continue increasing the backreaction, the two highlighted modes in Fig.~\ref{fig:QNMs_backreaction} remain purely imaginary until they collide (black dot). After the collision they have equal imaginary parts and opposite real parts. This agrees with the behavior of the non-hydrodynamic modes of the effective theory, which at $q = 0$ are the roots of $\omega^2 + 2iM_2\,\omega - M_1 = 0$. The coefficients can therefore be read off from the two modes as
\begin{equation}
    M_1 = -\,\omega_+\omega_-\ , \quad \quad \quad M_2 = \frac{i}{2}\left(\omega_+ + \omega_-\right) \ .
\end{equation}
The unstable mode becomes stable where $M_1$ changes sign (white dot). Since $\partial_\phi^2 V < 0$ on the spinodal branch, this is where $c_v$ vanishes. The collision takes place at $M_1 = M_2^2$ (black dot).

\section{Discussion}\label{sec:disc}

In this paper we have developed an effective field theory for first-order phase transitions, consisting of a real scalar order parameter coupled to a relativistic fluid. The EFT was constructed starting from an action for the scalar and an ideal fluid, with dissipative terms added at the level of the equations of motion. Linearizing around homogeneous equilibrium states yields the quartic dispersion relation \eqref{eq:dispersion}, whose four roots comprise two hydrodynamic (sound) modes and two gapped modes.

On the spinodal branch the second derivative of the effective potential for the scalar, $\partial_\phi^2 V$, is negative and the equilibrium state is dynamically unstable. The sign of the specific heat $c_v$ determines which mode carries the spinodal instability. Near the spinodal points bounding the unstable branch, $c_v$ is necessarily negative, and the instability resides in one of the hydrodynamic modes. The growth rate then vanishes at $q=0$ and peaks at finite momentum. Deeper inside the unstable branch, $c_v$ may cross zero and become positive. The hydrodynamic modes are then stable, and the instability is instead carried by one of the non-hydrodynamic modes, with a non-zero growth rate at $q=0$. In either case the band of unstable wavevectors is bounded in momentum by \eqref{eq:momentumdisp}. The scenario with $c_v>0$ on a large portion of the spinodal branch arises generically when the sector driving the transition is small compared to the full system, e.g.\ when $g_\text{eff}$ in \eqref{eq:weakpressure} is large.

We tested the EFT against a holographic model which, depending on the choice of parameters, exhibits spinodal branches with only hydrodynamic instabilities or with the instability switch. We fit the six parameter combinations entering \eqref{eq:dispersion} from the zero frequency value and the low momentum expansion of the dispersion relation of quasinormal modes. One of the parameters, the squared sound speed $c_s^2=s/c_v$, is independently computable from the holographic background. Another quantity that can be obtained directly from the background and enters in the QNM dispersion relations is the sound attenuation constant $\Gamma_s$. For both quantities, the background and QNM calculations agree within numerical precision.

With our fitted parameters, the EFT accurately reproduces the dispersion relation of the sound mode and the lightest non-hydrodynamic mode, in most cases up to $q\sim 2T$, and in some cases up to $q\sim 4T$. It typically outperforms pure hydrodynamics, particularly when there are collisions between the modes; these are captured by the EFT but completely absent in the hydrodynamics. As expected, the EFT is more accurate when the non-hydrodynamic modes are lighter than other gapped quasinormal modes present in the full microscopic theory, which usually happens near the spinodal points. In summary, the microscopic calculation confirms the validity of the EFT in the expected regime.

In the part of the spinodal branch where the specific heat is positive, the system is dynamically unstable but thermodynamically stable. This can be viewed as a counterexample to the so-called ``correlated stability conjecture'' (CSC) \cite{Gubser:2000mm}, which posits that a black brane with a non-compact horizon is classically dynamically stable if and only if it is locally thermodynamically stable. The conjecture works in one direction, a negative specific heat leads to unstable sound \cite{Buchel:2005nt}. Earlier counterexamples to the full conjecture were found in e.g. \cite{Friess:2005zp,Buchel:2010wk,Buchel:2020jfs}; these too involve different types of phase transitions. As explained in section~\ref{sec:hydrofluc}, one could treat the source for the scalar order parameter, $J$, as a generalized thermodynamic variable. In~\cite{Gouteraux:2024adm}, such generalization led to explaining the spontaneous condensation in holographic superfluid as a thermodynamical instability. On the other hand, the expectation value of the dual operator does not work in general as generalized thermodynamic variable \cite{Buchel:2011ra}. One should note that at equilibrium $J=0$, so non-zero $J$ either corresponds to deviations from equilibrium or a deformation of the theory, in contrast with the temperature and other ordinary thermodynamic variables. Neglecting this issue, the resulting susceptibility matrix \eqref{eq:suceptibility} always has exactly one unstable eigenvalue on the spinodal branch, matching the fact that there is always one dynamically unstable mode there. In this sense, allowing for a more general notion of thermodynamic variables, the CSC is valid here. 

One limitation of our construction concerns the field content of the EFT itself. The construction is justified by a separation of scales: near the spinodal the order-parameter mode becomes light, parametrically separated from the remaining tower of quasinormal modes, which are then integrated out. Away from the spinodal this separation degrades. As can be seen in for example Fig.~\ref{fig:gapped-QNMs-q=0}, near the metastable branches the gap of the order-parameter mode can grow until it crosses that of modes we have discarded, so that the EFT retains a heavy mode while omitting lighter ones. The ratio of this gap to that of the lightest excluded mode can thus provide a diagnostic for the regime of validity, with the description presumably becoming untrustworthy as the ratio approaches unity. Which of the excluded modes would need to be retained, and whether they admit a useful parametrization in the same spirit rather than simply reintroducing the full tower of modes, we leave as an open question.

In cosmological phase transitions, one of the most important as well as least accessible quantities is the terminal wall speed of expanding bubbles \cite{Hindmarsh:2020hop}. Standard determinations rely on a Boltzmann treatment of transport across the wall and therefore presuppose a weakly coupled plasma \cite{Moore:1995si}. In strongly coupled settings the wall speed has instead been obtained from direct numerical evolution in holographic theories \cite{Bea:2021zsu}. A natural next step would be to use the EFT we have constructed to predict the wall speed. The key player in this calculation is the parameter $\gamma(\phi,T)$, which gives rise to a friction force on the wall balancing the pressure difference that drives the expansion. Our EFT also differs from the usual case through the non-canonical kinetic terms $Z_0$ and $Z_s$, which can lead to interesting modifications.

It should be noted that our EFT is a truncated derivative expansion---and, as discussed above, a truncation in field content---whereas a bubble wall can involve large gradients. A computation of the wall speed is meaningful only to the extent that higher-derivative corrections remain negligible on the scale of the wall thickness. This is precisely what makes the comparison worth doing, since, for example, in \cite{Bea:2021zsu} the wall speed is read off from the nonlinear evolution of the full holographic theory. Agreement between the two would establish that the gradient expansion remains predictive far from equilibrium; disagreement could quantify where it breaks down. Extending the EFT to the full non-linear regime requires going beyond the analysis of the quasinormal mode spectrum, and additional subtleties may arise when determining the effective potential and kinetic terms, related to renormalization and higher order correlation functions \cite{Hoyos:2026eoq}. Despite these caveats, previous comparisons of critical bubbles from the EFT and the microscopic theory in holographic models suggest that the EFT could provide a good approximate description \cite{Henriksson:2025vci,Hoyos:2026eoq}.

Lastly, it would be interesting to follow the nonlinear evolution starting from a state on the spinodal branch where the specific heat is positive. Existing holographic simulations of spinodal decomposition begin from a region where the sound channel is unstable, so that the unstable direction is the conserved energy density and the system is driven directly towards phase separation \cite{Attems:2017ezz,Janik:2017ykj,Attems:2019yqn}. For spinodal states with stable sound, the growth rate still peaks at nonzero wavenumber, but the instability now extends down to $q=0$. For a predominantly homogeneous initial perturbation, one would expect an initial stage of essentially homogeneous relaxation, with the released energy transferred to the thermal sector through the dissipative coupling. Whether the system is driven towards phase separation in the end depends on a competition: faster-growing inhomogeneities may become nonlinear before the homogeneous relaxation has carried the system out of the spinodal region. If they do not, and the total energy lies within the coexistence range, the system may instead settle in a metastable homogeneous state, from which phase separation would require nucleation. If the total energy lies outside the coexistence range, the final state should be a stable homogeneous state with the same energy, possibly reached after a transient formation and disappearance of structure at the unstable wavelengths.

\section*{Acknowledgments}

We would like to thank Blaise Gout\'eraux and Elias Kiritsis for useful comments. A.C. acknowledges support from the German Research Foundation (DFG) through the Collaborative Research Center ToCoTronics, Project-ID 258499086--SFB 1170, and from Germany's Excellence Strategy through the W{\"u}rzburg-Dresden Cluster of Excellence on Complexity, Topology and Dynamics in Quantum Matter - ctd.qmat (EXC 2147, Project-ID 390858490). 
O.H. acknowledges support from the Waldemar von Frenckell foundation. 
C.H. acknowledges support from the Spanish Agencia Estatal de Investigaci\'on and Ministerio de Ciencia, Innovaci\'on y Universidades through the grant PID2024-161500NB-I00.
M.S.G. acknowledges support from a Royal Society - Research Ireland University Research Fellowship via grants URF/R1/211027 and RF/ERE/231191.
P.T. acknowledges support from the project ``Dark Energy and the Origin of the Universe'' (PRE2022-102220) funded by MCIN/AEI/10.13039/501100011033, from grant No.~00017375 from the Simons Foundation, and from project PID2022-141125NB-I00 from the Spanish Ministry of Science, Innovation and Universities.

\section*{AI usage}
The \texttt{Julia} package \texttt{QUASTELL}~\cite{QUASTELL} was developed by the authors based on a pre-existing \texttt{Mathematica} code, which was originally written without the use of AI tools. The AI coding assistant Claude Code was used only to debug and clean up the \texttt{Julia} code and repository, and Claude and ChatGPT were used to clean up the \texttt{Mathematica} notebooks in the repository. Claude was also used to revise the manuscript and polish the language. AI tools did not contribute to the scientific content of this work, and the authors take full responsibility for the code, results and manuscript.

\appendix

\section{Expectation values of dual operators}\label{app:holoren}

In order to obtain the gauge theory expectation values as a function of these fall-off coefficients, we can change coordinate to Fefferman-Graham (FG),
\begin{equation}
    ds^2 = L^2\left(\gamma_{\mu\nu}dx^\mu dx^\nu+ \frac{du^2}{u^2}\right),
\end{equation}
with $\gamma_{\mu u} = 0$, and use the known relation between the fall-offs in such coordinates and the expectation values, see e.g.~\cite{Attems:2017zam}, 
\begin{equation}
\begin{aligned}
    \langle T_{\mu\nu}\rangle = & \frac{L^3}{4\pi G_5}\left(\gamma_{\mu\nu}^{(4)}+\gamma^{(0)}_{\mu\nu}\left(\psi_{(0)}\psi_{(2)}-\frac{\psi_{(0)}^4}{18}-\frac{\psi_{(0)}^4}{4\psi_M^2}\right)\right),\\
    \langle \mathcal{O}\rangle = & \frac{L^3}{4\pi G_5}\left(-2\psi_{(2)}+\frac{\psi_{(0)}^3}{\psi_M^2}\right),
\end{aligned}
\end{equation}
where,
\begin{equation}
\begin{aligned}
    \gamma_{\mu\nu} & = \frac{1}{u^2}\eta_{\mu\nu} + \gamma_{\mu\nu}^{(2)} + \gamma_{\mu\nu}^{(4)}u^2+\mathcal{O}\left(u^4\right),\\
    \psi & = \psi_{(0)} u +\psi_{(2)} u^3 + \mathcal{O}\left(u^5\right),
\end{aligned}
\end{equation}
with $\eta_{\mu\nu}$ the boundary, Minkowski metric. Near the boundary, our coordinate $r$ relates to the FG $u$ as,
\begin{equation}
    r = u + \frac{f_4}{8}u^5 + \mathcal{O}\left(u^7\right),
\end{equation}
which we use to read-off all the FG fall-off coefficients in terms of ours,
\begin{equation}
    \begin{aligned}
        \psi & = \psi_s u + \psi_2 u^3+\mathcal{O}\left(u^5\right),\\
        \gamma_{tt} & = -\frac{f(r) g(r)}{r^2} = -\frac{1}{u^2} + \frac{\psi_s^2}{3}+\left(-\frac34 f_4+\frac{1}{2}\psi_s\psi_2-\frac{\psi_s^4}{18}\right)u^2+\mathcal{O}\left(u^4\right),\\
        \gamma_{ii} & = \frac{g(r)}{r^2} =\frac{1}{u^2} - \frac{\psi_s^2}{3}+\left(-\frac14 f_4-\frac{1}{2}\psi_s\psi_2+\frac{\psi_s^4}{18}\right)u^2+\mathcal{O}\left(u^4\right).\\
    \end{aligned}
\end{equation}
Upon substitution of the coefficients in the expectation value expressions, we arrive at
\begin{equation}
    \begin{aligned}
        \varepsilon \equiv & \frac{4\pi G_5}{L^3}\left\langle T^{tt}\right\rangle = -\frac34 f_4 - \frac12 \psi_s\psi_2 + \frac{\psi_s^4}{4\psi_M^2},\\
        p \equiv & \frac{4\pi G_5}{L^3}\left\langle T^{ii}\right\rangle = -\frac14 f_4 + \frac12 \psi_s\psi_2 - \frac{\psi_s^4}{4\psi_M^2},\\
        \phi \equiv & \frac{4\pi G_5}{L^3}\left\langle \mathcal{O}_{\psi}\right\rangle = -2\psi_2 + \frac{\psi_s^3}{\psi_M^2},\\
    \end{aligned}
    \label{eq:VEVs}
\end{equation}
where $\varepsilon$, $p$ and $\phi$ are the energy density, pressure and scalar expectation value respectively. The Ward identity reads,
\begin{equation}
     3p-\varepsilon = -\psi_s \phi,
\end{equation}
where $\psi_s$ manifests itself as the conformal symmetry breaking mass scale of the boundary theory.

\section{EFT parameters for theory B}
\label{app:TheoryBplots}

\begin{figure}[t!]
    \centering
    \includegraphics[width=0.48\linewidth]{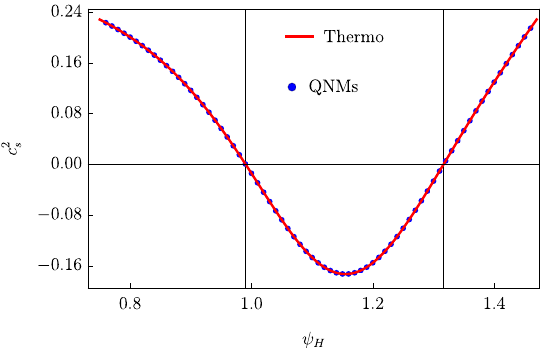}
    \hfill
    \includegraphics[width=0.49\linewidth]{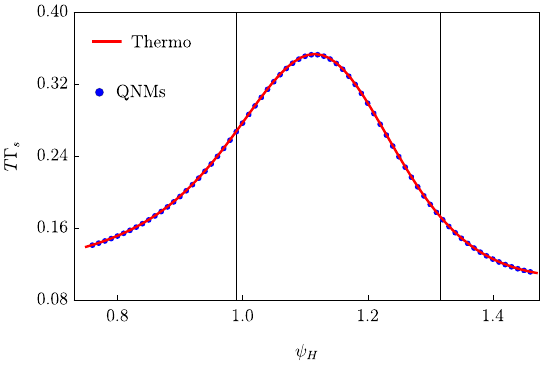}
    
    \caption{Speed of sound squared (left) and sound attenuation constant $\Gamma_s$ (right) as a function of the scalar field at the horizon for theory B. The red lines are obtained from the background solutions using the relations (\ref{eq:cs2fromthermo}) (left) and (\ref{eq:GammaS}) (right), whereas the blue dots are computed from the sound QNMs, namely (\ref{eq:alphasound1}) (left) and (\ref{eq:alphasound2}) (right). Vertical black lines represent the boundary of the spinodal branch.}
    \label{fig:cs2andGammascomparisonTheoryB}
\end{figure}

\begin{figure}[t!]
    \centering
    \includegraphics[width=0.55\linewidth]{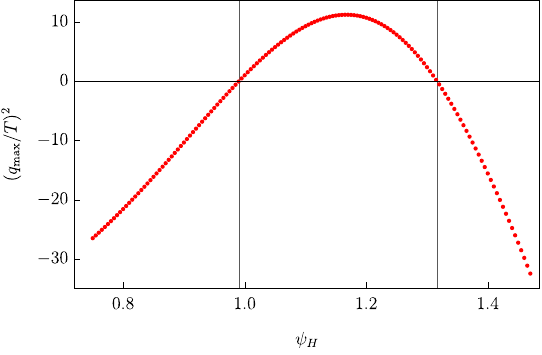}
    \caption{Value of $(q_{\text{max}}/T)^2$ as a function of the scalar field at the horizon $\psi_H$ for theory B. Vertical black lines represent the boundary of the spinodal branch.}
    \label{fig:qmaxtheoryB}
\end{figure}

\begin{figure}[t!]
    \centering
  
  \includegraphics[width=0.46\linewidth]{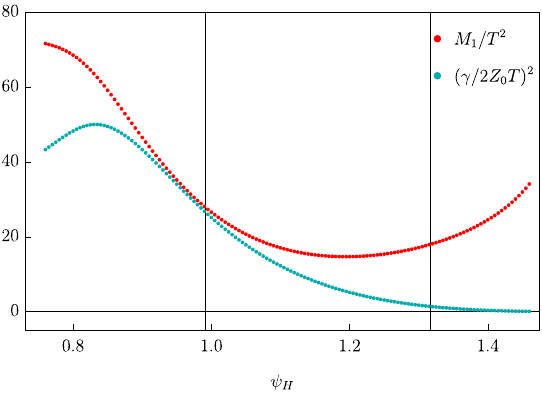}
  \hspace{2mm}  \includegraphics[width=0.495\linewidth]{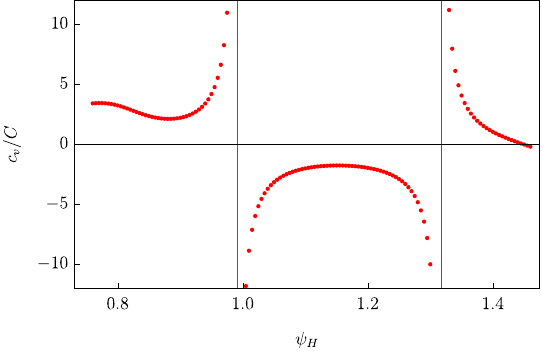}\hfill
    \includegraphics[width=0.47\linewidth]{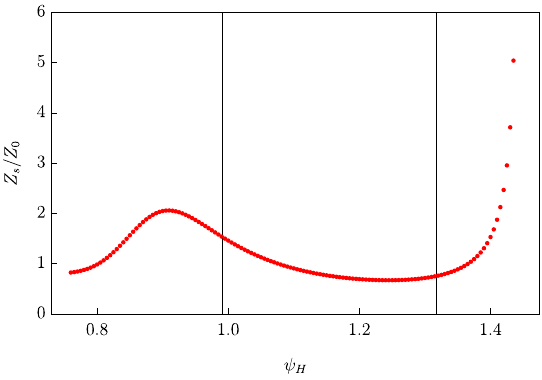}\hfill
    \includegraphics[width=0.49\linewidth]{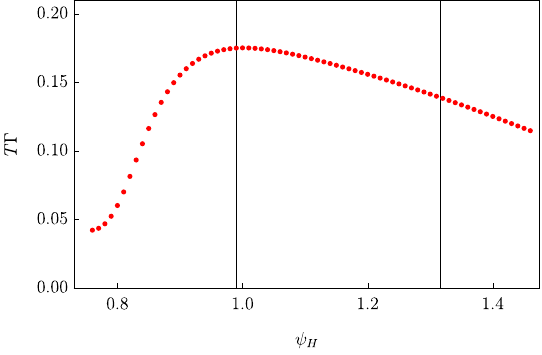}\hfill
    \caption{ Effective theory parameters as functions of the scalar field at the horizon for theory B: $M_1$ and $M_2^2 = \left( \gamma/2 Z_0\right) ^2$  (top-left), $c_v/C$ (top-right), $Z_s/Z_0$  (bottom-left) and $\Gamma$ (bottom-right). All dimensionful quantities are expressed in units of temperature.   Vertical black lines represent the boundary of the spinodal branch.}
    \label{fig:EFT-parametersTheoryB}
\end{figure}

In this appendix, we collect the results of the effective field theory parameters (\ref{eq: EFT parameters}) for theory B, corresponding to $\psi_M=1$, whose phase diagram is provided in the right plot of Fig.~\ref{fig:s-of-T}. The EFT parameters are obtained following the steps outlined in section~\ref{sec:extractingEFTcoeffs}.
In theory B, the two gapped modes always display a non-vanishing real part and never collide, as can be seen in Fig.~\ref{fig:gapped-QNMs-q=0-theoryB}. This means that $\Delta = \sqrt{M_1-M_2^2}$ is always real and never vanishes. In this theory, the instability in the spinodal region is always due to the sound mode. As a result, the only special points are the spinodal ones.

In Fig.~\ref{fig:cs2andGammascomparisonTheoryB}, we show the speed of sound squared (left plot) and the sound attenuation constant (right plot). These quantities can be computed both from the background data and from the QNMs. The plots display the results using both computations, showing good agreement within numerical precision. In Fig.~\ref{fig:qmaxtheoryB}, we show the zero-frequency mode at finite momentum $q_{\text{max}}$. It is positive in the spinodal branch and negative in the locally stable branches, consistent with expectations from (\ref{eq:momentumdisp}).

In Fig.~\ref{fig:EFT-parametersTheoryB}, we show the remaining EFT parameters. There, we
can see that $M_1$ is always greater than the square of $M_2=\gamma/2 Z_0$, consistent with the fact that $\Delta$ is always real and never vanishes. Moreover, $c_v/C$ is always negative in the spinodal region and positive outside, except for some points at large $\psi_H$. Correspondingly, $Z_s/Z_0$ blows up. The same pattern is also observed for theory A close to the first point where the gapped modes collide.

\bibliographystyle{JHEP}
\bibliography{biblio}

\end{document}